\documentclass[12pt]{article}
\usepackage{geometry}
\usepackage{graphicx} 
\usepackage{amsmath,amssymb,amsthm, gensymb}
\usepackage{hyperref}
\usepackage{microtype}
\usepackage{booktabs, float}
\usepackage{color}
\usepackage{svg}
\usepackage{diagbox}
\usepackage{multirow}
\usepackage{tabularx}
\newcolumntype{Y}{>{\centering\arraybackslash}X}
\DeclareMathOperator{\avg}{avg}
\usepackage{xurl}

\begin{document}

\title{How geometry of subduction zones correlates with earthquake dynamics}

\author{Oscar Y. L. Chau$^{1}$, Rebecca Bendick$^{2}$, Gary P. T. Choi$^{1,\ast}$, L. Mahadevan$^{3,4,\ast}$ \\
\\
\footnotesize{$^1$Department of Mathematics, The Chinese University of Hong Kong, Hong Kong, China}\\
\footnotesize{$^2$EarthScope Consortium, Washington, D.C., USA}\\
\footnotesize{$^3$School of Engineering and Applied Sciences, Harvard University, Cambridge, MA, USA}\\
\footnotesize{$^4$Departments of Physics, and Organismic and Evolutionary Biology, Harvard University, Cambridge, MA, USA}\\
\footnotesize{$^\ast$To whom correspondence should be addressed; E-mail: ptchoi@cuhk.edu.hk, lmahadev@g.harvard.edu}}
\date{ }

\maketitle

\begin{abstract}
Subduction zones on the surface of the Earth, where abrupt sliding leads to earthquakes, are generally curved and localized. How does the geometry of these zones influence the occurrence of megathrust earthquakes? Here we use a combination of simple scaling arguments and data analysis using the differential geometry of surfaces to examine the relationship between the earthquake productivity of subduction zones and their shape. A scaling argument  suggests how interface curvature changes both the accumulation and release of stress relative to planar interfaces; conformable sliding along relatively flat subduction zones should lead to rare but large events, while curved subduction zones should lead to frequent smaller events. To test this, we leverage global geometry datasets and analyze the correlation between the surface curvatures of the subduction zones and the frequency and magnitude of earthquakes therein. Our analysis shows that weakly curved slab geometries are associated with rarer larger magnitude events, while slab geometries with a larger relative dispersion in curvature are associated with frequent but smaller magnitude events. Using different scale-dependent shape metrics of the subduction zones, we show that the earthquake productivity is influenced by the conformability of the overriding and downgoing plates. More broadly, our results suggest the need to incorporate the large-scale geometry of subduction zones in computational models and predictive frameworks for earthquake risk.  
\end{abstract}

\section*{Introduction}
Megaquakes, events with moment magnitudes greater than 8.0, are rare but contribute disproportionately to fatalities and economic losses due to earthquakes and tsunamis.  All known megaquakes except two poorly understood Portuguese events have occurred on subduction interfaces, including the recent (29 July 2025) Petropavlovsk-Kamchatsky event. A natural question is how the nature of subduction zones affects moment release or the spatio-temporal distribution of seismicity in these regions, informing both earthquake dynamics and characterization of time-dependent megaquake risk.\\

Subduction zones are defined as the interface between two domains: a downgoing plate (lower plate) and an overriding plate (upper plate).  Over the past several decades, there have been a wide range of studies about the geometry of subducted plates~\cite{laravie1975geometry,bayly1982geometry}, the relationship between curvature and plate rigidity~\cite{bevis1986curvature}, physical models of subduction zone curvature~\cite{mahadevan2010subduction}, extreme localized exhumation~\cite{bendick2014extreme}, and the geometric control of earthquake rupture~\cite{plescia2020geometric} and megathrust earthquake cycle~\cite{biemiller2024subduction}.  However, these typically consider the geometry of only the downgoing subduction slab, such as for geometric stiffening, rather than the contribution of the interaction of two domains on the subduction interface.  Interfacial contributions to moment budgets have mainly been restricted to studies of the role of roughness on coupling~\cite{lambert2025competition}.\\

\begin{figure}[t!]
    \centering
    \includegraphics[width=\linewidth]{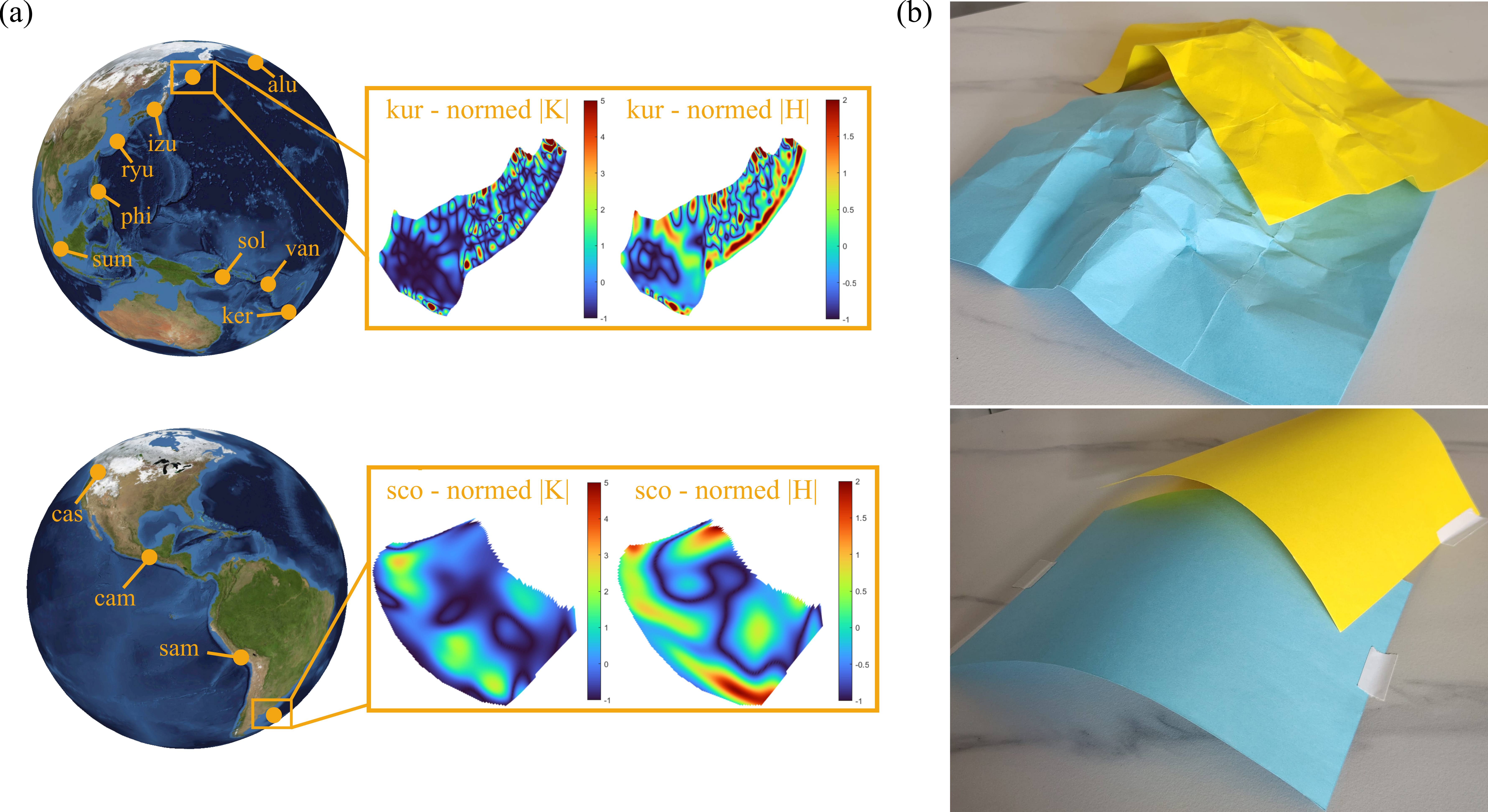}
    \caption{\textbf{Subduction zone models and their surface curvatures.} (a)~The locations of the 13 subduction zone models considered in this study: Aleutians (alu), Central America (cam), Cascadia (cas), Izu-Bonin (izu), Kermadec (ker), Kuril (kur), Philippines (phi), Ryukyu (ryu), South America (sam), Scotia (sco), Solomon Islands (sol), Sumatra/Java (sum), and Vanuatu (van), with a zoom-in at kur and sco color-coded with the normalized absolute Gaussian curvature ($|K| -$mean($|K|$))/mean($|K|$) and normalized absolute mean curvature ($|H| -$mean($|H|$))/mean($|H|$) for the illustration. {(b)~Paper models of rough (top) and smooth (bottom) subduction surfaces. In each image, the blue and yellow paper models represent the down-going plate and over-riding plate respectively.}}
    
    \label{fig:F1_illustration}
\end{figure}

Our ability to characterize the three-dimensional shape of subduction zone geometry models has recently become possible due to increased knowledge from tomographic and seismic imaging, such as the Slab1.0 model~\cite{hayes2012slab1,pagani2014openquake,bletery2016mega}. More recently, Hayes et al.~\cite{hayes2018slab2} developed the Slab2 model with improved 3D geometries of all known active subduction zones globally, thereby allowing for a more detailed analysis of the geometry of subduction zones. Generalized characterization of geometric variation of the subduction zones using Slab2 provides along-strike and along-dip curvatures. While these directional curvatures offer a geologically intuitive decomposition of surface bending and are commonly used to interpret subduction zone morphology and variations in tectonic stress, based on a common cylindrical approximation of subduction zones, the typical directional curvatures are inherently direction-dependent and do not fully characterize the intrinsic geometry of a surface, even as the intrinsic geometry likely contributes to the mechanics of a complex interface.\\

In contrast, geometric contributions to surface interactions between two objects with relative velocity are a common topic of investigation throughout materials science, occurring wherever solids move past one another.  As a result, there is an extensive literature describing both the influence of the frictional characteristics of the interface, such as roughness~\cite{sagy2007evolution,candela2012roughness,brodsky2016constraints}, and the influence of geometric characteristics of the interface, such as curvature~\cite{mahadevan2010subduction}. A classic example of this is the Lake--Thomas effect, in which interfaces exhibit hysteresis in energy budgets due to the cost of unpeeling and re-adhering on a nonplanar interface~\cite{memet2021static}. Here we consider a variant of this effect on geological scales, and show that this allows us to explain the magnitude and frequency of earthquakes on non-planar subduction interfaces.\\

As a start, we utilize differential geometry and shape representation methods to study the geometry of subduction zones and its relationship with the subduction zone earthquake productivity, and relate this to elementary notions borrowed from materials science. Specifically, we consider the Slab2 dataset~\cite{hayes2018slab2} and focus on 13 subduction zone models including Aleutians (alu), Central America (cam), Cascadia (cas), Izu-Bonin (izu), Kermadec (ker), Kuril (kur), Philippines (phi), Ryukyu (ryu), South America (sam), Scotia (sco), Solomon Islands (sol), Sumatra/Java (sum), and Vanuatu (van) (see Fig.~\ref{fig:F1_illustration}(a)). Instead of directional curvatures, we use the surface curvatures of subduction zone geometry models to capture their geometric features. By analyzing the correlation between different measurements of these surface curvatures and the frequency and magnitude of earthquakes for different subduction zones from the historical data, we show that the earthquake productivity depends heavily on the magnitude of the curvatures but not on their sign. As illustrated in Fig.~\ref{fig:F1_illustration}, some subduction zones are generally rougher (e.g., kur) while others are flatter (e.g., sco). Moreover, we combine the surface curvature quantification approach with geometric simplification methods to analyze the correlation between earthquake productivity and the slab geometry at different characteristic scales, yielding key insights into the importance of conformability in seismic productivity.

\section*{Methods}

\subsection*{Curvature-controlled rupture scaling}
On geological scales, the mechanics of megathrust earthquakes are governed by frictional slip on a curved interface between two deformable plates. This curvature imposes geometric constraints on the kinematic compatibility of slip. On a planar interface, slip over a large region can occur coherently with relatively uniform stress drop. In contrast, on a curved interface, finite slip across an extended patch induces variations in slip direction, normal stress, and in-plane deformation of the surrounding lithosphere. These effects introduce an additional curvature-dependent elastic penalty that grows with rupture size and limits the spatial extent of coherent slip.

A minimal scaling argument can be constructed by comparing elastic energy release with this geometric penalty. For a rupture patch of size $L$ and slip $\delta$, the released elastic energy scales as $E_{\mathrm{rel}} \sim \Delta\tau\,\delta\,L^2$, where $\Delta\tau$ is the stress drop. Let $\kappa_{\mathrm{eff}}$ denote a characteristic magnitude of surface curvature, defined in terms of the mean and Gaussian curvatures, e.g. $\kappa_{\mathrm{eff}} \sim |H| \quad \text{or} \quad \kappa_{\mathrm{eff}} \sim \sqrt{|K|}.$ Curvature induces a geometric incompatibility that generates additional elastic strain of order $\delta\,\kappa_{\mathrm{eff}}$.  We expect that increasing curvature magnitude reduces the maximum spatial extent of coherent rupture owing to the incompatibility of the surfaces undergoing relative slip. To see this, we note that the coherent rupture length $L_c$ may be understood as the maximum scale over which slip can occur with approximately uniform stress drop and kinematic compatibility. On a curved interface, a rupture patch of size $L$ experiences a rotation of the local tangent frame of order $\Delta \theta \sim \kappa_{\mathrm{eff}} L$. A slip displacement $\delta$ across the patch therefore generates a mismatch displacement $\delta_{\mathrm{mis}} \sim \delta \kappa_{\mathrm{eff}} L$, corresponding to an effective strain $\varepsilon_{\mathrm{geom}} \sim \delta \kappa_{\mathrm{eff}}$ in the surrounding lithosphere. The resulting geometric stress perturbation scales as $\Delta\sigma_{\mathrm{geom}} \sim Y \delta \kappa_{\mathrm{eff}}^2 L$, where $Y$ is an effective in-plane modulus.   The associated elastic energy penalty therefore scales as $E_{\mathrm{geom}} \sim Y \delta^2 \kappa_{\mathrm{eff}}^2 {L^3}$.
Comparing this with the elastic energy release yields a curvature-limited coherent rupture length 
\begin{equation}
L_c \sim \frac{\Delta\tau}{Y\,\delta \kappa_{\mathrm{eff}}^2}.
\end{equation}
This expression shows that increasing curvature magnitude reduces the maximum spatial extent of coherent rupture by introducing a super-extensive energetic penalty associated with geometric incompatibility. The above geometric constraint translates directly into scaling laws for earthquake magnitude and frequency. Since the maximum seismic moment scales as
\begin{equation}
M_{0,\max} \sim \mu\,\delta\,L_c^2 \sim \frac{\mu (\Delta \tau)^2}{Y^2 \delta \kappa_{\mathrm{eff}}^2},
\end{equation}
this implies a power-law decrease of maximum moment with curvature magnitude. At the same time, for a given tectonic loading rate, smaller rupture sizes imply shorter recurrence intervals, so that event frequency increases with curvature. This simple argument suggests that curvature acts as a control parameter for rupture segmentation: highly curved subduction zones favor smaller, more frequent events, while weakly curved and more conformable interfaces permit larger, less frequent earthquakes. To test this, we now turn to a careful analysis of the data on surface curvature and earthquake productivity.

\subsection*{Slab geometry quantification using surface curvature}
To capture the key geometric features of the slab surfaces, we consider the \emph{Gaussian curvature} and the \emph{mean curvature}, two fundamental concepts in differential geometry~\cite{do2016differential}  (see also SI Fig.~S1). \\

Mathematically, the Gaussian curvature is defined at each point on a smooth surface as the product of the principal curvatures: $K = \kappa_1 \kappa_2$, where \( \kappa_1 \) and \( \kappa_2 \) denote the maximum and minimum normal curvatures at the point. Note that the Gaussian curvature is \emph{intrinsic}, meaning that it is invariant under coordinate transformations and depends solely on the surface's internal geometry. It classifies points locally: \( K > 0 \) corresponds to elliptic points, \( K < 0 \) to hyperbolic points, and \( K = 0 \) to parabolic or planar regions. Complementing the Gaussian curvature, the mean curvature provides another intrinsic measure of surface geometry, capturing a different aspect of shape. It is defined at each point on a smooth surface as the average of the principal curvatures: $H = \frac{1}{2}(\kappa_1 + \kappa_2)$, where $\kappa_1$ and $\kappa_2$ are the maximum and minimum normal curvatures at the point. Unlike Gaussian curvature, which reflects how a surface bends in orthogonal directions simultaneously, mean curvature quantifies the average bending and is particularly useful in detecting regions of overall convexity or concavity. Importantly, it is also an \textit{extrinsic} quantity---its value depends on how the surface is embedded in space, making it sensitive to global deformation while still providing local geometric insight.\\

Now, given a subduction zone geometry model represented by data points with (depth, latitude, longitude) = $(d, \phi, \lambda)$ in a gridded form, we first obtain the Cartesian coordinates using
\begin{equation}
(X,Y,Z) = (R \cos \phi \cos \lambda,   R \cos \phi \sin \lambda, R \sin \phi),
\end{equation}
where $R = 6371 + d$. Then, the Gaussian curvature and mean curvature can be calculated using
\begin{equation} \label{eqt:K}
K = \frac{LN - M^2}{EG - F^2}
\end{equation}
and
\begin{equation} \label{eqt:H}
H = \frac{EN+GL-2FM}{2(EG-F^2)},
\end{equation}
where $\begin{pmatrix} E & F\\F & G\end{pmatrix}$ and $\begin{pmatrix} L & M\\M & N\end{pmatrix}$ are the first fundamental form and second fundamental form. In practice, the above formulas can be easily discretized using finite differences based on the grid data points.\\

It is easy to see that the mean and Gaussian curvatures $H$, $K$ are defined everywhere on the slab surfaces and can easily capture the geometric features in different regions. As the curvature values at different positions of the surface may have opposite signs, one may also consider the absolute mean and Gaussian curvatures $|H|, |K|$ as well as the normalized versions of them to focus on their magnitudes (see SI Fig.~S2--S7 for the geometries of the 13 subduction zone models color-coded with the surface curvatures). \\

Besides, the geometric quantification based on surface curvatures can be easily combined with shape simplification procedures to analyze the subduction zone geometry at different characteristic scales. More specifically, instead of taking all grid points in a slab geometry model, we consider a space separation parameter $s$ and extract the grid points with a spacing of $s$ along each grid direction, which results in a simplified surface representation. As the simplified surface representation remains to be in a gridded data form, we can repeat the calculation of the surface curvatures as described in the above section. 

\section*{Results}

\subsection*{Correlation between surface curvature and earthquake productivity}
To study the relationship between the geometry of subduction zones and the earthquake productivity, we compute the Gaussian curvature $K$ and mean curvature $H$ at each point of the surface reconstructed from the depth, latitude, and longitude data from the Slab2 dataset~\cite{hayes2018slab2}. Note that the surface curvatures $K, H$ and the absolute surface curvatures $|K|, |H|$ vary spatially over the entire slab geometry model. To further quantify them as a scalar value for studying their relationship with the number of occurrences and maximum magnitude of earthquakes, we consider the minimum, maximum, and average values of them, as well as their standard deviation and coefficient of variation, from which we can easily calculate the Pearson correlation coefficient for each geometric quantity and the earthquake productivity for the 13 slab models:
\begin{equation}
    \rho_{X,Y} = \frac{\text{cov}(X,Y)}{\sigma_X \sigma_Y},
\end{equation}
where $X$ is a geometric quantity (minimum, maximum, mean, standard deviation (SD), and coefficient of variation (CV) of each of $H$, $K$, $|H|$, $|K|$) and $Y$ is an earthquake productivity measurement (the maximum magnitude $M_{\text{max}}$, average magnitude $M_{\text{avg}}$, number of earthquakes $n$), $\text{cov}(\cdot, \cdot)$ is the covariance, and $\sigma_X, \sigma_Y$ are the standard deviation of $X$ and $Y$ respectively. Here, the data of earthquake magnitudes and numbers are obtained from the USGS Earthquake Catalog (\url{https://earthquake.usgs.gov/earthquakes/search/}), with the period of time covered being from 1900-01-01 to 2025-12-31 (see SI Fig.~S8--S9 for histograms of the earthquake events). To simplify our discussion, in the main text we focus on the earthquakes with magnitude threshold $M_w \geq 6.0$. See SI Section S2 for the analysis on the earthquake data with different magnitude thresholds ($M_w \geq 5.0, 5.5, 6.0, 6.5, 7.0$).\\

\begin{table}[t!]
    \centering
    \begin{tabularx}{\textwidth}{l|Y|Y|Y}
        \toprule
        \diagbox[width=2cm]{$X$}{$Y$}
        & \textbf{Maximum Magnitude $M_{\max}$} & \textbf{Average Magnitude $M_{\avg}$} & \textbf{Number of Earthquakes $n$}\\
        \hline
        \hline
        Mean($|H|$) & \textbf{-0.5911} & \textbf{-0.5937} & -0.2226 \\ 
        Mean($|K|$) & \textbf{-0.6037} & \textbf{-0.5979} & -0.2758 \\
        SD($|H|$)&	\textbf{-0.6444}&	\textbf{-0.6581}&	-0.2026\\
        SD($|K|$)&	\textbf{-0.6023}&	\textbf{-0.6274}&	-0.1462\\
        CV($|H|$)&	0.0322&	-0.0187&	0.2908\\
        CV($|K|$)&	0.0773&	-0.0193&	\textbf{0.5276}\\
        \bottomrule
    \end{tabularx}
    \caption{\textbf{The correlation coefficient $\rho_{X,Y}$ between the surface curvature quantities $X$ and earthquake productivity measurements $Y$ of subduction zones.} For each geometric quantity (average, standard deviation, and coefficient of variation of the absolute mean curvature $|H|$ and absolute Gaussian curvature $|K|$), we calculate the Pearson Correlation coefficients between it and the maximum historically observed magnitude $M_{\max}$, the average magnitude $M_{\avg}$, and the number of earthquakes $n$ for 13 slab geometry models from the Slab2 dataset with magnitude threshold $M_w \geq 6.0$. Correlation coefficients with absolute value $\geq 0.5$ are in boldface. See also SI Table S1 for a more complete correlation table.}
    \label{table:statistics1}
\end{table}

\begin{figure*}[t!]
    \centering
    \includegraphics[width=\textwidth]{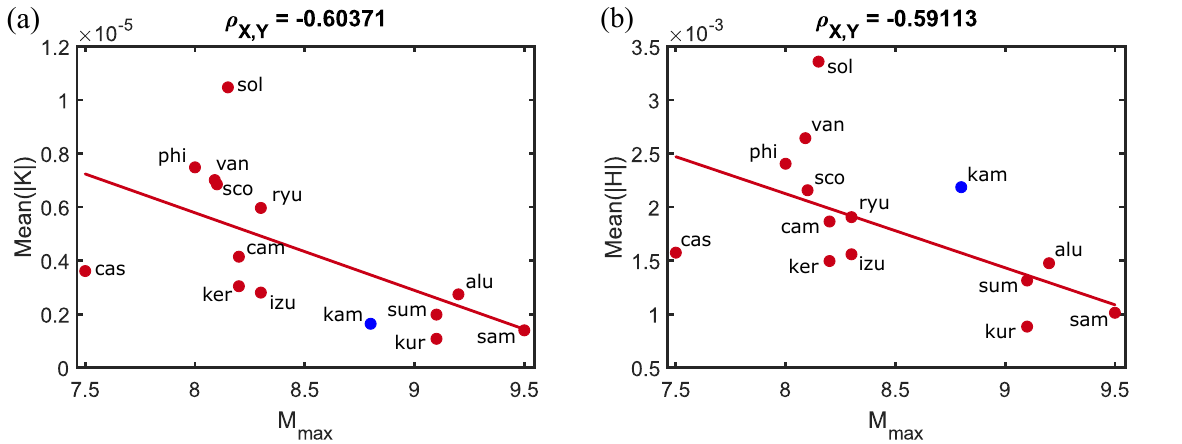}
    \caption{\textbf{Correlation between surface curvature and the maximum historically observed magnitude of earthquakes $M_{\max}$.} (a)~The average absolute Gaussian curvature mean($|K|$). (b)~The average absolute mean curvature mean($|H|$). Each point represents one slab model, and the red line is the best-fit straight line. The new Petropavlovsk-Kamchatsky event (kam) is highlighted in blue. See the caption of Fig.~\ref{fig:F1_illustration} for the full name of each model.}
    \label{fig:curvature_vs_mag}
\end{figure*}

\begin{figure*}[t!]
    \centering
    \includegraphics[width=0.5\textwidth]{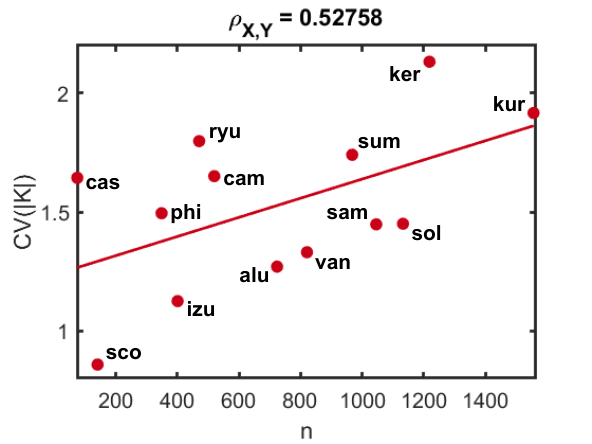}
    \caption{\textbf{Correlation between coefficient of variation of $|K|$ and the number of occurrence of earthquakes.} Each point represents one slab model, and the red line is the best-fit straight line. See the caption of Fig.~\ref{fig:F1_illustration} for the full name of each model.}
    \label{fig:CVabsK_num}
\end{figure*}

As shown in Table~\ref{table:statistics1}, both mean($|H|$) and mean($|K|$) show a notable negative correlation with both the maximum historically observed magnitude of earthquakes and average magnitude of earthquakes (Fig.~\ref{fig:curvature_vs_mag}). Also, both the standard deviation of $|H|$ and that of $|K|$ show a strong negative correlation with the average magnitude of earthquakes. By contrast, the coefficient of variation of absolute Gaussian curvature (CV($|K|$)) shows a notable positive correlation with the number of earthquakes (Fig.~\ref{fig:CVabsK_num}). Overall, this indicates that weakly curved subduction zones are associated with rare larger earthquakes, while those with greater curvature variation are associated with more frequent and smaller earthquakes. See SI Table~S1 for a complete correlation table between more surface curvature quantities and earthquake productivity measurements, and SI Fig.~S10--S12 for more correlation plots.

Besides, comparing the correlation coefficients associated with $H, K$ and those with $|H|, |K|$ in SI Table~S1 (see also SI Table~S2--S3 for other magnitude thresholds), we can see that the absolute values of the curvatures are more predictive than the signed curvatures. We note that the Solomon Islands (sol) subduction zone appears as an outlier. This supports our hypothesis that subduction interface geometry and earthquake productivity are related when they are coupled through tectonic dynamics.  Currently, the geometry of the Solomon slab (see also SI Fig.~S1--S4) does not represent the tectonic dynamics because the polarity of that subduction zone and the plate kinematics have changed in the past 10 Ma~\cite{taylor2024oceanic}, thus the curvatures are not representative of the stress accumulation and release in the current earthquake cycle.  \\

In Fig.~\ref{fig:curvature_vs_mag}, we further include the 29 July 2025 Petropavlovsk-Kamchatsky event (kam, with a maximum magnitude of 8.8) and the corresponding geometric measures, from which we can see that the event is consistent with our observed correlation. In other words, the result shows that the earthquake productivity depends on the overall magnitude of the curvatures but not their sign. This supports our interpretation that the conformability of the two domains across the subduction interface influences the earthquake energy budget.  If the geometric stiffness of one or the other domains (rather than the interface itself) was the important characteristic, we would expect negative curvatures (concave shapes) to be relatively soft, thus weak and positive curvatures (convex shapes) to be relatively stiff and therefore for the sign of the curvature to be highly correlated to productivity.\\

All together, the observed correlations between the magnitude of the surface curvatures and earthquake productivity are consistent with the curvature-controlled rupture scaling interpretation, and with the finding that the absolute values of the mean and Gaussian curvatures are more predictive than their signs. It suggests that geometric incompatibility, rather than the distinction between convex and concave regions, governs the energy budget and rupture coherence on subduction interfaces.

\subsection*{Analysis with depth uncertainty}
Also, in the Slab2 model~\cite{hayes2018slab2}, due to instrument resolution and environmental variation, there is a certain amount of uncertainty when measuring the depth of each point on the surface. In particular, the Slab2 model assumed the vertical location of a data point to be described by a normal distribution, where the mean is the preferred depth $d$ and the standard deviation is the uncertainty $u$. Therefore, we also incorporate the uncertainty information into our geometric analysis. Instead of directly using the depth $d$ in the curvature calculation, we consider the modified depth $\tilde{d} = d + k u$, where $u$ is the given uncertainty and $k$ is a scalar factor. By setting different values of $k$, we can effectively add or subtract different levels of uncertainty and create multiple new surfaces (see SI Section~S3 and Fig.~S13 for more details). Then, we can perform the correlation analysis between the surface curvature based on the modified depth and the maximum magnitude, average magnitude, and number of earthquakes. From the correlation results  (see SI Table~S4--S10 and Fig.~S14--S16), a consistent trend with our previous analysis can be observed. This suggests that our findings of the correlation between surface curvature and earthquake productivity are robust to the uncertainty in the depth measurement.

\subsection*{Analyzing different characteristic scales}
In the Slab2 dataset, each grid data point is measured with a space separation of 5.5km ($\sim0.05^{\circ}$ in longitude or latitude). To analyze different characteristic scales, we can further consider space separations of 16.5km, 27.5km, 38.5km, and 49.5km and compute the curvatures of each simplified surface (see SI Section~S4 and SI Fig.~S17--S18 for more details). Then, for each choice of the space separation, we can repeat the correlation analysis as in the previous section (see SI Table~S10--S12 and Fig.~S19), from which we can see that the average absolute mean and Gaussian curvatures remain negatively correlated with the both the maximum historically observed magnitude $M_{\max}$ and the average magnitude $M_{\text{avg}}$. These consistent negative correlations across different characteristic scales suggest that conformability is more important than roughness for large earthquake events.

\section*{Discussion}
In this work, we have presented a new approach for studying the subduction zone geometry and the earthquake productivity using different geometric tools and geometric simplifications. By examining different quantities derived from the surface curvature calculation, we have found that the magnitude of both the mean curvature $H$ and the Gaussian curvature $K$ is most negatively correlated with earthquake magnitude, and the relative dispersion in curvature is positively correlated with earthquake productivity determined by the number of catalog events in the subduction segment.  These correlations are valid even after taking the uncertainty of the depth measurement into consideration. Moreover, by considering different space separations, we have shown that the correlation between surface curvature and earthquake magnitude of subduction zones remains high at different characteristic scales, suggesting that conformability is more important than roughness. Altogether, our work offers a new framework for understanding the subduction zone geometry and earthquake events.\\

This result has implications for understanding the mechanics of stick-slip energy budgets on nonplanar interfaces at scales of $O(1000\text{km})$, extending the scale invariance of geometric effects by several orders of magnitude.  It also has practical implications for the characterization of seismic hazard in subduction domains in the absence of complete earthquake catalogs, which may take thousands of years to accumulate.  For example, the correlation of less curved interfaces, both concave and convex, to larger maximum-moment events and lower earthquake productivity implies that such areas may have less frequent but larger and more destructive earthquakes, while highly curved interfaces may slip more often with smaller maximum magnitudes.  At least, the true geometry of subduction interfaces and consequent geometric effects on the energy (moment) budgets must be incorporated into subduction zone mechanics, rather than simply approximating these systems as planar or cylindrical in form.

\bibliographystyle{ieeetr}
\bibliography{reference}

\begin{thebibliography}{10}

\bibitem{laravie1975geometry}
J.~A. Laravie, ``Geometry and lateral strain of subducted plates in island arcs,'' {\em Geology}, vol.~3, no.~9, pp.~484--486, 1975.

\bibitem{bayly1982geometry}
B.~Bayly, ``Geometry of subducted plates and island arcs viewed as a buckling problem,'' {\em Geology}, vol.~10, no.~12, pp.~629--632, 1982.

\bibitem{bevis1986curvature}
M.~Bevis, ``The curvature of {W}adati-{B}enioff zones and the torsional rigidity of subducting plates,'' {\em Nature}, vol.~323, no.~6083, pp.~52--53, 1986.

\bibitem{mahadevan2010subduction}
L.~Mahadevan, R.~Bendick, and H.~Liang, ``Why subduction zones are curved,'' {\em Tectonics}, vol.~29, no.~6, 2010.

\bibitem{bendick2014extreme}
R.~Bendick and T.~A. Ehlers, ``Extreme localized exhumation at syntaxes initiated by subduction geometry,'' {\em Geophys. Res. Lett.}, vol.~41, no.~16, pp.~5861--5867, 2014.

\bibitem{plescia2020geometric}
S.~M. Plescia and G.~P. Hayes, ``Geometric controls on megathrust earthquakes,'' {\em Geophys. J. Int.}, vol.~222, no.~2, pp.~1270--1282, 2020.

\bibitem{biemiller2024subduction}
J.~Biemiller, A.-A. Gabriel, D.~A. May, and L.~Staisch, ``Subduction zone geometry modulates the megathrust earthquake cycle: Magnitude, recurrence, and variability,'' {\em J. Geophys. Res. Solid Earth}, vol.~129, no.~8, p.~e2024JB029191, 2024.

\bibitem{lambert2025competition}
V.~Lambert and E.~E. Brodsky, ``Competition between roughness and strength for scale-dependent surfaces,'' {\em Phys. Rev. E.}, vol.~111, no.~6, p.~065502, 2025.

\bibitem{hayes2012slab1}
G.~P. Hayes, D.~J. Wald, and R.~L. Johnson, ``Slab1.0: A three-dimensional model of global subduction zone geometries,'' {\em J. Geophys. Res. Solid Earth}, vol.~117, no.~B1, 2012.

\bibitem{pagani2014openquake}
M.~Pagani, D.~Monelli, G.~Weatherill, L.~Danciu, H.~Crowley, V.~Silva, P.~Henshaw, L.~Butler, M.~Nastasi, L.~Panzeri, {\em et~al.}, ``{OpenQuake} engine: An open hazard (and risk) software for the global earthquake model,'' {\em Seismol. Res. Lett.}, vol.~85, no.~3, pp.~692--702, 2014.

\bibitem{bletery2016mega}
Q.~Bletery, A.~M. Thomas, A.~W. Rempel, L.~Karlstrom, A.~Sladen, and L.~De~Barros, ``Mega-earthquakes rupture flat megathrusts,'' {\em Science}, vol.~354, no.~6315, pp.~1027--1031, 2016.

\bibitem{hayes2018slab2}
G.~P. Hayes, G.~L. Moore, D.~E. Portner, M.~Hearne, H.~Flamme, M.~Furtney, and G.~M. Smoczyk, ``Slab2, a comprehensive subduction zone geometry model,'' {\em Science}, vol.~362, no.~6410, pp.~58--61, 2018.

\bibitem{sagy2007evolution}
A.~Sagy, E.~E. Brodsky, and G.~J. Axen, ``Evolution of fault-surface roughness with slip,'' {\em Geology}, vol.~35, no.~3, pp.~283--286, 2007.

\bibitem{candela2012roughness}
T.~Candela, F.~Renard, Y.~Klinger, K.~Mair, J.~Schmittbuhl, and E.~E. Brodsky, ``Roughness of fault surfaces over nine decades of length scales,'' {\em J. Geophys. Res. Solid Earth}, vol.~117, no.~B8, 2012.

\bibitem{brodsky2016constraints}
E.~E. Brodsky, J.~D. Kirkpatrick, and T.~Candela, ``Constraints from fault roughness on the scale-dependent strength of rocks,'' {\em Geology}, vol.~44, no.~1, pp.~19--22, 2016.

\bibitem{memet2021static}
E.~Memet, F.~Hilitski, Z.~Dogic, and L.~Mahadevan, ``Static adhesion hysteresis in elastic structures,'' {\em Soft Matter}, vol.~17, no.~10, pp.~2704--2710, 2021.

\bibitem{do2016differential}
M.~P. Do~Carmo, {\em Differential geometry of curves and surfaces: revised and updated second edition}.
\newblock Dover Publications, Garden City, NY, 2016.

\bibitem{taylor2024oceanic}
B.~Taylor and E.~K. Benyshek, ``Oceanic plateau and spreading ridge subduction accompanying arc reversal in the {S}olomon {I}slands,'' {\em Geochem. Geophys. Geosyst.}, vol.~25, no.~1, p.~e2023GC011270, 2024.

\end{thebibliography}

\clearpage

\centerline{\Large\textbf{Supplementary Information}}
\appendix
\renewcommand\thefigure{S\arabic{figure}}    
\setcounter{figure}{0}
\renewcommand\thetable{S\arabic{table}}    
\setcounter{table}{0}
\renewcommand{\thesection}{S\arabic{section}}


\section{Slab surface geometry}

\begin{figure}[b!]
    \centering
    \includegraphics[width=\linewidth]{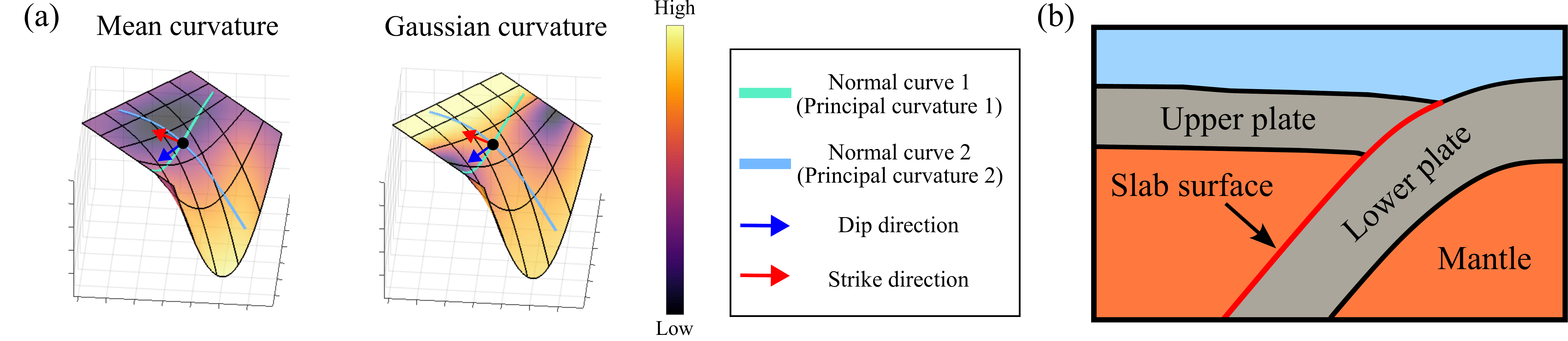}
    \caption{\textbf{Subduction zone and curvature illustrations.} (a)~An illustration of the mean curvature $H$ and the Gaussian curvature $K$ of a surface with an arbitrary point highlighted. Both the mean and Gaussian curvature at the point can be expressed using the principal curvatures $\kappa_1, \kappa_2$, with $H = (\kappa_1+\kappa_2)/2$ and $K = \kappa_1 \kappa_2$. It is noteworthy that the principal curvature directions are not necessarily identical to the conventional dip and strike directions. (b)~An illustration of subduction interfaces.}
    \label{fig:SF1_illustration}
\end{figure}

In this work, we analyzed the slab surface geometry of different subduction zones using surface curvatures. See Fig.~\ref{fig:SF1_illustration} for an illustration.

\begin{figure*}[t!]
    \centering
    \includegraphics[width=\textwidth]{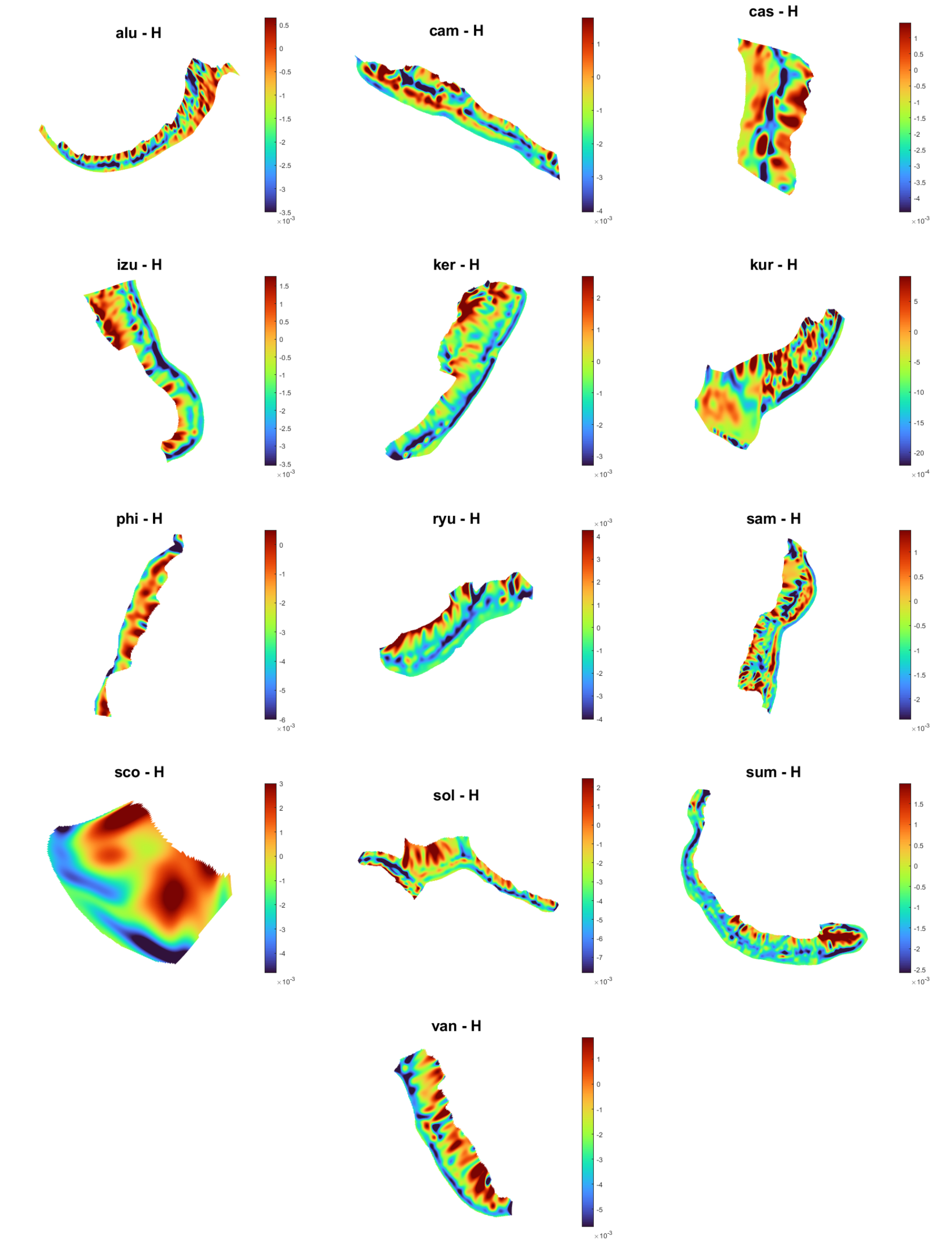}
    \caption{\textbf{Visualization of the mean curvature $H$ in each of the 13 subduction zones.}}
    \label{fig:all_H_plots}
\end{figure*}

\begin{figure*}[t!]
    \centering
    \includegraphics[width=\textwidth]{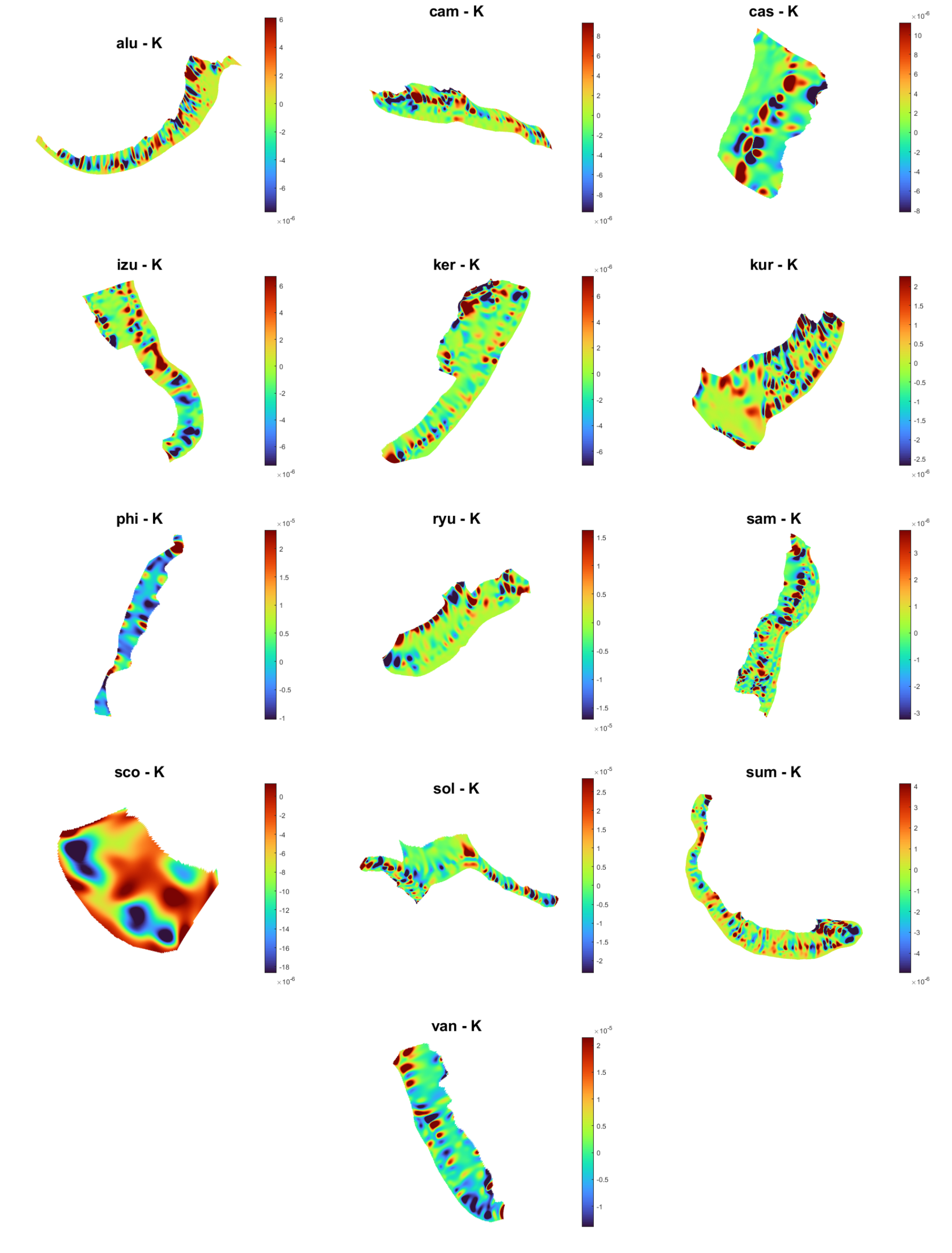}
    \caption{\textbf{Visualization of the Gaussian curvature $K$ in each of the 13 subduction zones.}}
    \label{fig:all_K_plots}
\end{figure*}

\begin{figure*}[t!]
    \centering
    \includegraphics[width=\textwidth]{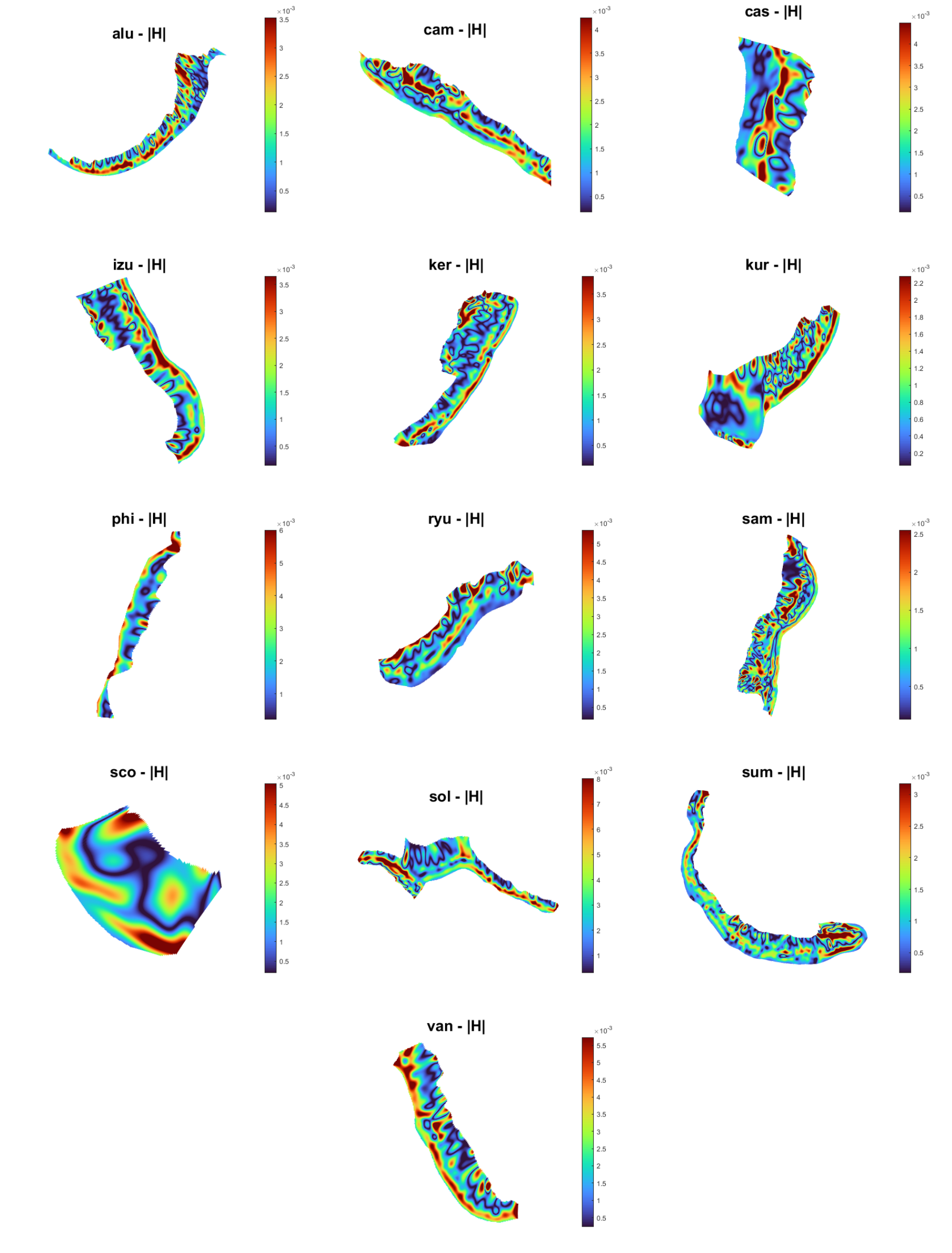}
    \caption{\textbf{Visualization of the absolute mean curvature $|H|$ in each of the 13 subduction zones.}}
    \label{fig:all_absH_plots}
\end{figure*}

\begin{figure*}[t!]
    \centering
    \includegraphics[width=\textwidth]{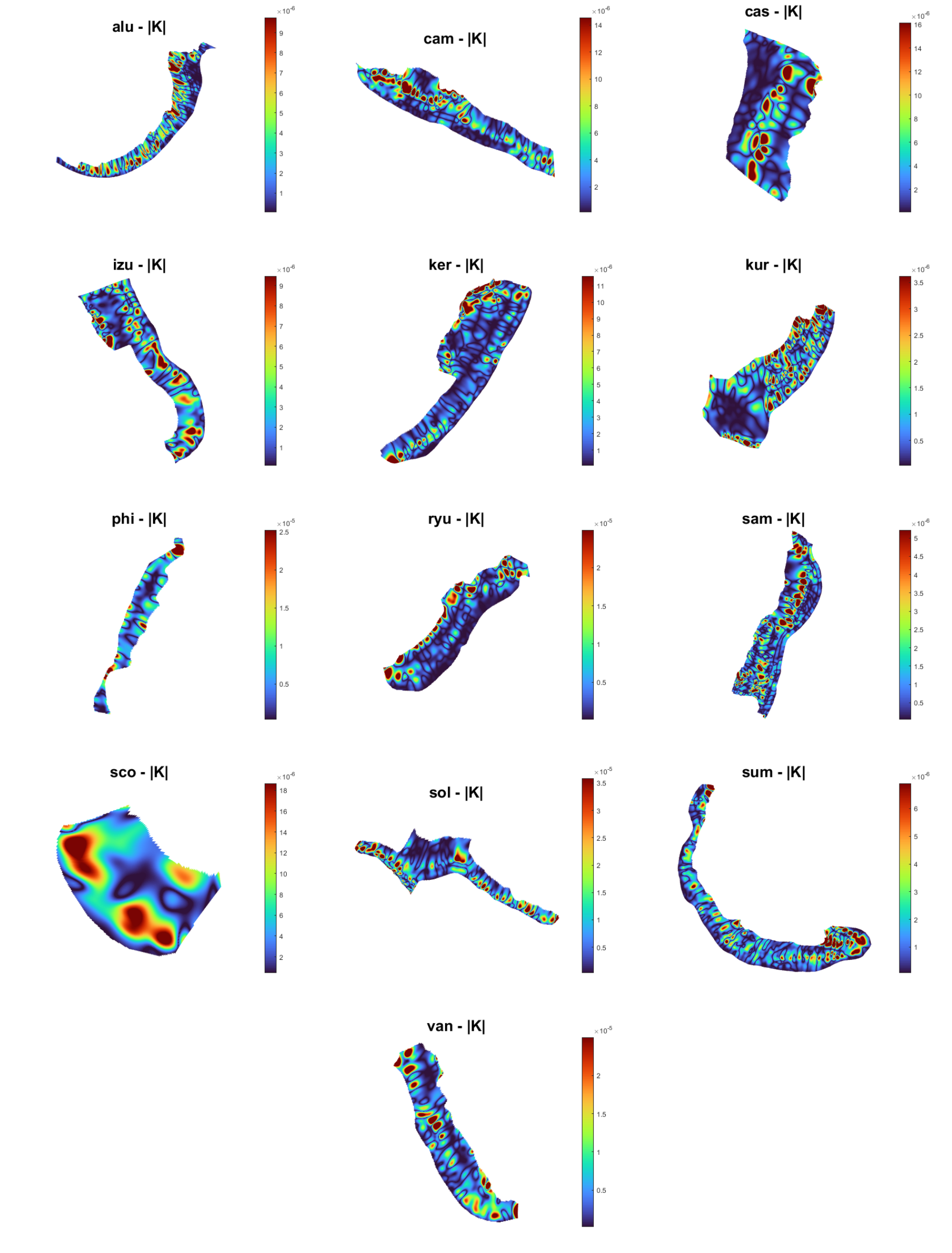}
    \caption{\textbf{Visualization of the absolute Gaussian curvature $|K|$ in each of the 13 subduction zones.}}
    \label{fig:all_absK_plots}
\end{figure*}

\begin{figure*}[t!]
    \centering
    \includegraphics[width=\textwidth]{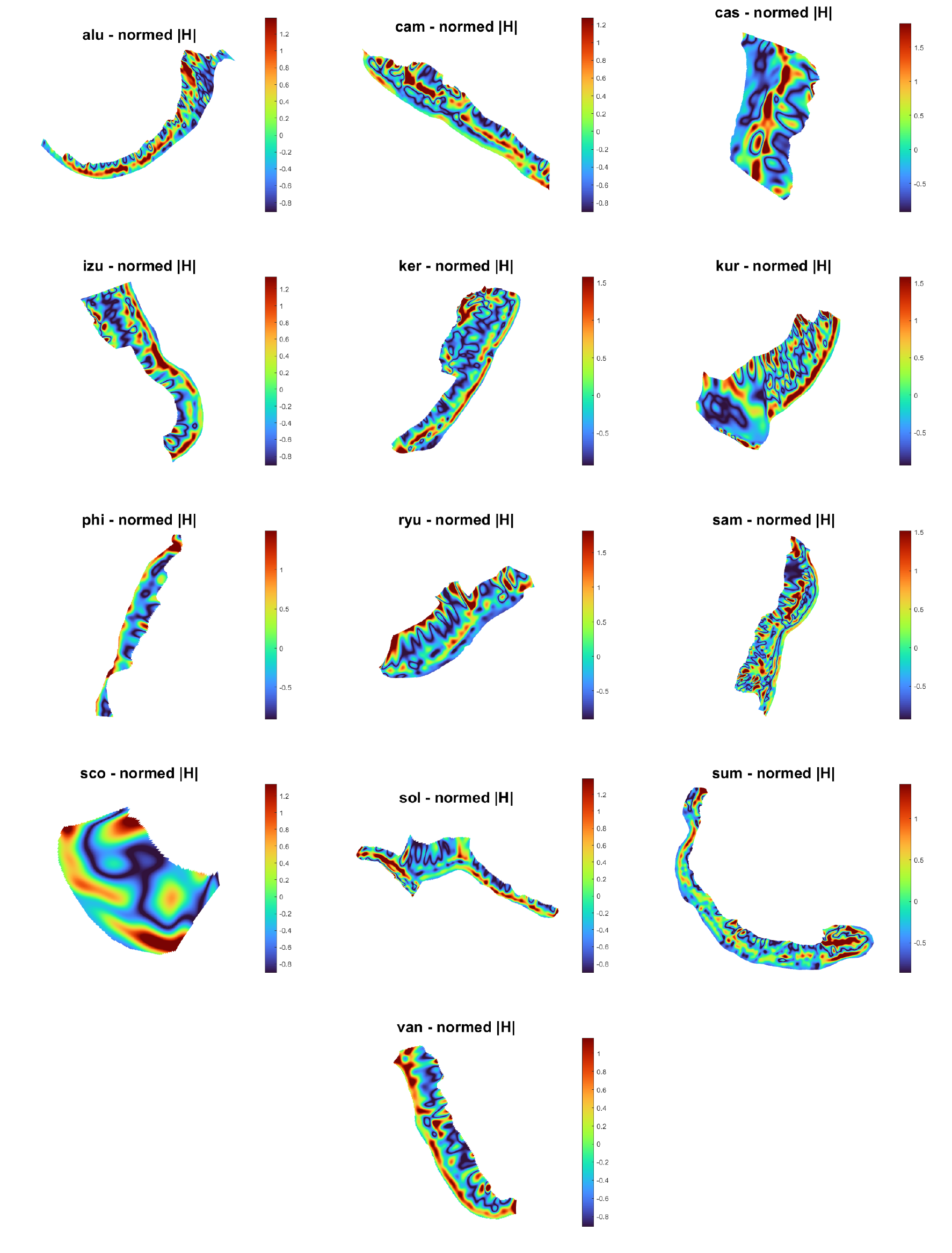}
    \caption{\textbf{Visualization of the normalized absolute mean curvature ($|H|-$mean($|H|$))/mean($|H|$) in each of the 13 subduction zones.}}
    \label{fig:all_normedabsH_plots}
\end{figure*}

\begin{figure*}[t!]
    \centering
    \includegraphics[width=\textwidth]{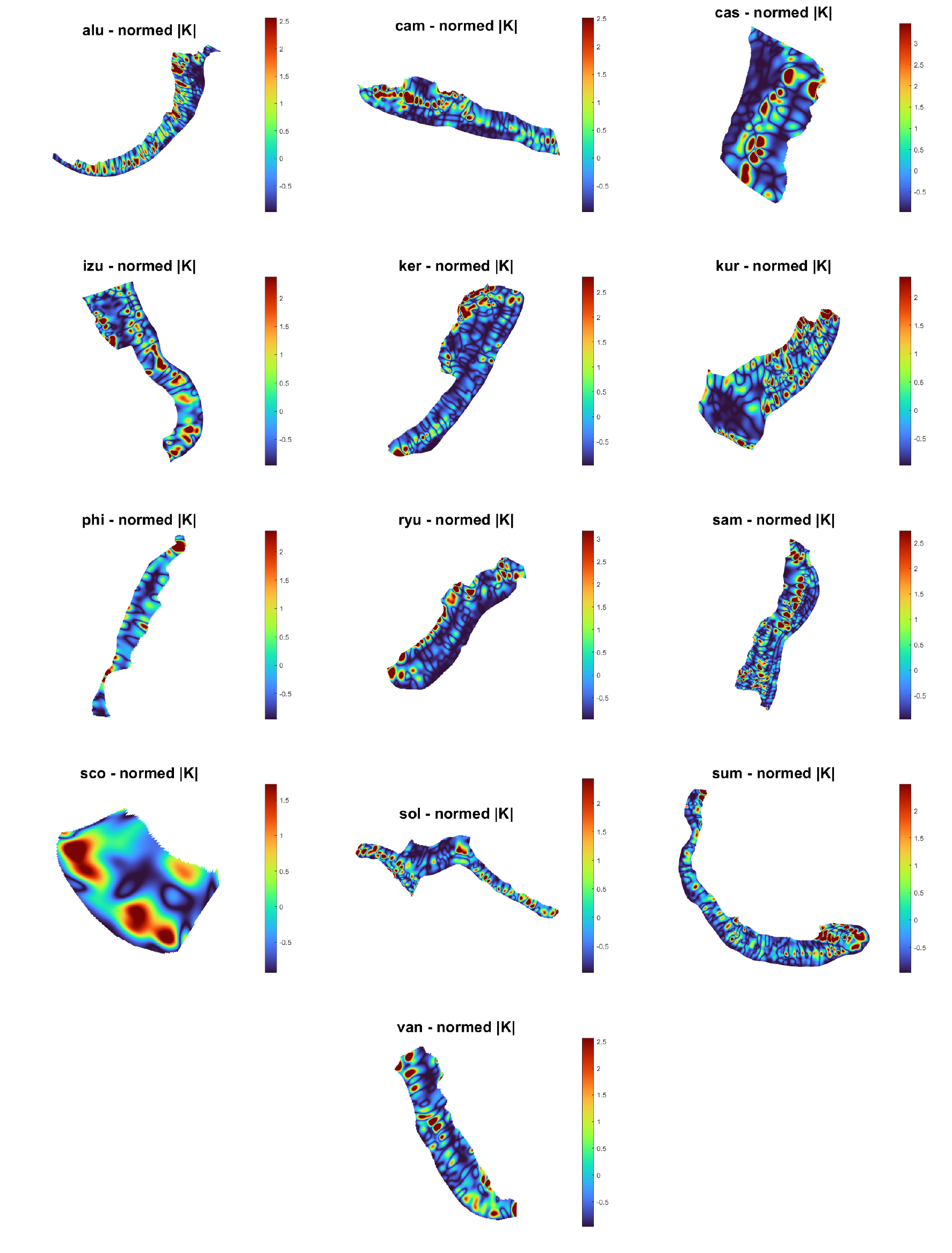}
    \caption{\textbf{Visualization of the normalized absolute Gaussian curvature ($|K|-$mean($|K|$))/mean($|K|$) in each of the 13 subduction zones.}}
    \label{fig:all_normedabsK_plots}
\end{figure*}

Besides directly considering the mean curvature $H$ and the Gaussian curvature $K$ of the slab surfaces, one can also focus on their magnitude and consider $|H|$ and $|K|$. Plots of the 13 subduction zones in the Slab2 dataset considered in this work (Aleutians (alu), Central America (cam), Cascadia (cas), Izu-Bonin (izu), Kermadec (ker), Kuril (kur), Philippines (phi), Ryukyu (ryu), South America (sam), Scotia (sco), Solomon Islands (sol), Sumatra/Java (sum), and Vanuatu (van)), color-coded with the mean curvature $H$, Gaussian curvature $K$, absolute mean curvature $|H|$, absolute Gaussian curvature $|K|$ and their normalized versions, are presented in Fig.~\ref{fig:all_H_plots}--\ref{fig:all_normedabsK_plots}.

\section{Analysis on earthquakes with different magnitude thresholds}

\begin{figure*}[t!]
    \centering
    \includegraphics[width=\textwidth]{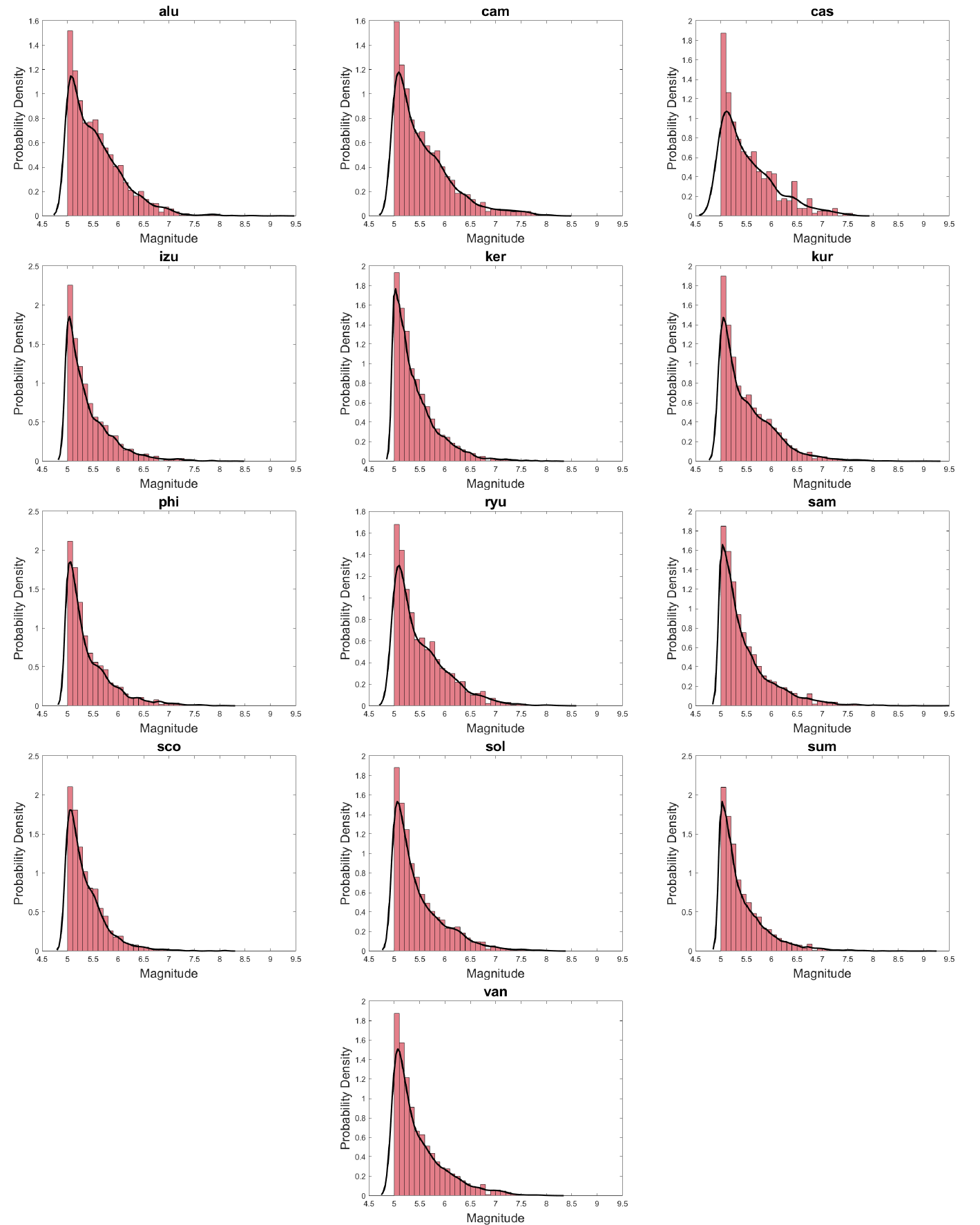}
    \caption{\textbf{Histogram and probability density estimation of the earthquake events in each of the 13 subduction zones.}}
    \label{fig:all_pdf_plots}
\end{figure*}

\begin{figure*}[t!]
    \centering
    \includegraphics[width=0.7\textwidth]{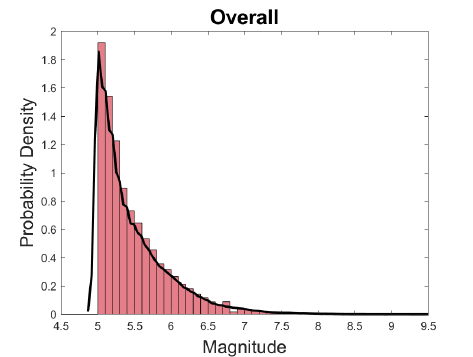}
    \caption{\textbf{Histogram and probability density estimation of total counts of earthquake events in all 13 subduction zones.}}
    \label{fig:overall_pdf_plots}
\end{figure*}

In the main text, we focus on the earthquake data with the magnitude threshold $M_w \geq 6.0$ for the calculation of the average magnitude $M_{\text{avg}}$ and number of earthquakes $n$, from which we further perform correlation analysis with the slab surface geometry. In Fig.~\ref{fig:all_pdf_plots}, we plot the histogram and probability density estimation of the earthquake events from 1900-01-01 to 2025-12-31 recorded in the USGS Earthquake Catalog for each of the 13 subduction zones. In Fig.~\ref{fig:overall_pdf_plots}, we summarize the events in all 13 subduction zones. \\

\begin{figure*}[t!]
    \centering
    \includegraphics[width=\textwidth]{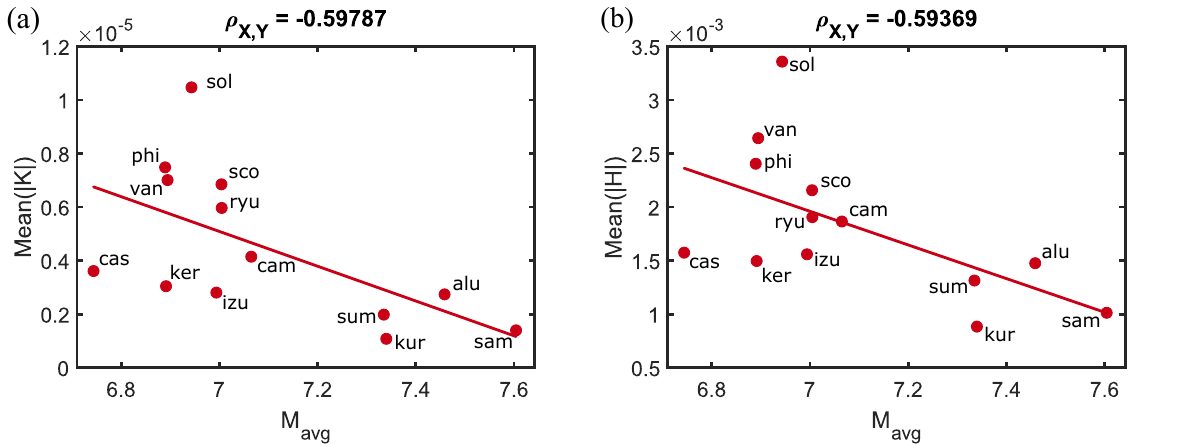}
    \caption{\textbf{Correlation between surface curvature and the average magnitude of earthquakes $M_{\text{avg}}$.} (a)~The average absolute Gaussian curvature mean($|K|$). (b)~The average absolute mean curvature mean($|H|$). Here, the average magnitude data is obtained based on the magnitude threshold $M_w \geq 6.0$. Each point represents one slab model, and the red line is the best-fit straight line. See the caption of main text Fig.~1 for the full name of each model.} 
    \label{fig:curvature_vs_avgmag}
\end{figure*}

\begin{figure*}[t!]
    \centering
    \includegraphics[width=\textwidth]{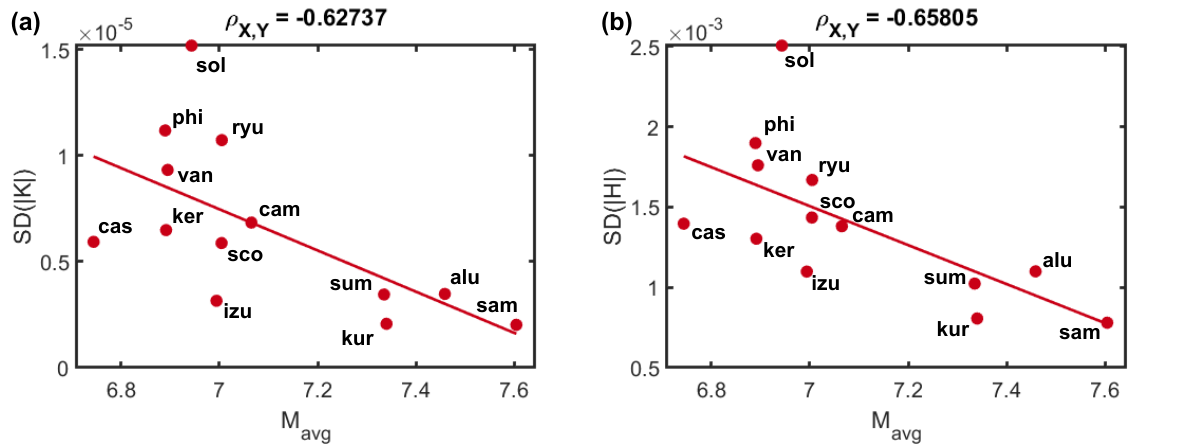}
    \caption{\textbf{Correlation between standard deviation of curvatures and the average magnitude of earthquakes.} (a)~The standard deviation of the Gaussian curvature SD($|K|$). (b)~The standard deviation of the mean curvature SD($|H|$). Each point represents one slab model, and the red line is the best-fit straight line. See the caption of main text Fig.~1 for the full name of each model.}
    \label{fig:SDabs_Mavg}
\end{figure*}

A complete correlation table is presented in Table~\ref{tab:SI_full_statistics}. Besides the results highlighted in the main text, in Fig.~\ref{fig:curvature_vs_avgmag} we consider the correlation between surface curvature and the average magnitude of
earthquakes $M_{\text{avg}}$. It can be observed that analogous to the maximum historically observed magnitude of earthquakes $M_{\max}$, here the average magnitude $M_{\text{avg}}$ shows a notable correlation with both the average values of $|K|$ and $|H|$. In Fig.~\ref{fig:SDabs_Mavg}, we also see a similar trend between the standard deviations of $|H|$, $|K|$ and $M_{\text{avg}}$. However, note that the correlation coefficients for the standard deviation are highly different from those for the coefficient of variation. A possible explanation is that the standard deviation quantities are unnormalized and therefore strongly influenced by the magnitude of the original curvature values. By contrast, after removing the effect of the magnitude, we can see that there is no linear relationship between the earthquake magnitude and the coefficient of variations $\text{CV}(|H|)$, $\text{CV}(|K|)$.\\

\begin{figure*}[t!]
    \centering
    \includegraphics[width=\textwidth]{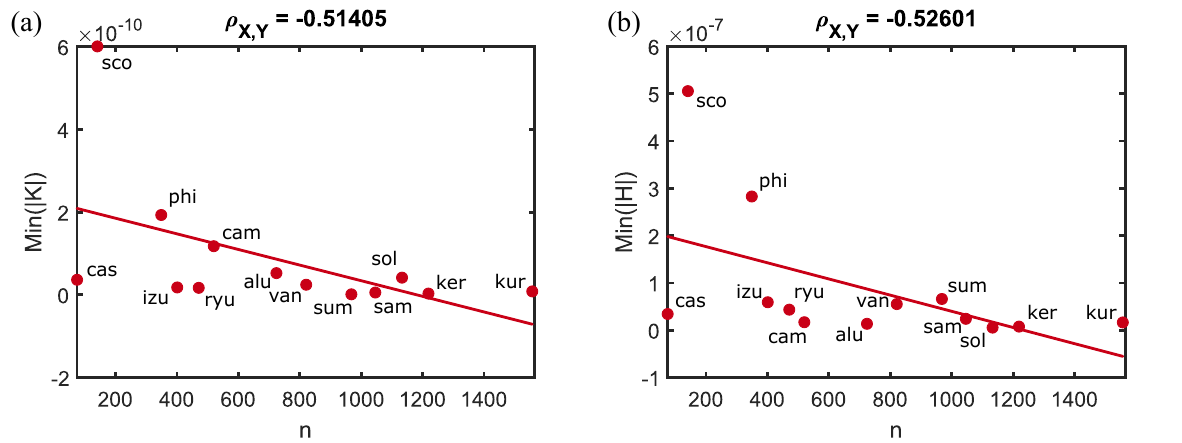}
    \caption{\textbf{Correlation between surface curvature and the number of earthquakes $n$.} (a)~The minimum absolute Gaussian curvature min($|K|$). (b)~The minimum absolute mean curvature min($|H|$). Here, the earthquake numbers are obtained based on the magnitude threshold $M_w \geq 6.0$. Each point represents one slab model, and the red line is the best-fit straight line. See the caption of main text Fig.~1 for the full name of each model.}
    \label{fig:curvature_vs_number}
\end{figure*}

Note that besides the correlation pairs shown in the main text, one can also see from the correlation analysis that there is a notable negative correlation ($\rho_{X,Y}\approx -0.5$) between the minimum absolute mean and Gaussian curvature (min$(|K|)$, min$(|H|)$) and the number of earthquakes $n$. However, as shown in Fig.~\ref{fig:curvature_vs_number}, such correlation coefficients are likely biased by the outlier (Solomon Islands) and hence are less indicative.\\

Besides, it is natural to ask how the correlation changes with the choice of the earthquake magnitude threshold. Here, we present additional results with different magnitude thresholds. In Table~\ref{tab:SI_avgmag}, we report the correlation coefficients between the mean and Gaussian curvatures and the average earthquake magnitude $M_{\text{avg}}$ calculated from data points with different magnitude thresholds $M_w \geq 5.0,  5.5,  6.0,  6.5,  7.0$. It can be observed that there is a notable negative correlation between the average absolute mean and Gaussian curvatures $\text{mean}(|H|), \text{mean}(|K|)$ and the average earthquake magnitude $M_{\text{avg}}$ regardless of the magnitude thresholds. In Table~\ref{tab:SI_num}, we present the correlation coefficients between the mean and Gaussian curvatures and the number of earthquakes $n$, again calculated from data points with different magnitude thresholds $M_w \geq 5.0,  5.5,  6.0,  6.5,  7.0$. We note that the correlation coefficients between the minimum absolute mean and Gaussian curvatures $\text{min}(|H|), \text{min}(|K|)$ and the number of earthquakes $n$ vary largely with the magnitude thresholds. This again suggests that the minimum absolute curvature values are less indicative.

\section{Surface geometry with depth uncertainty}

As described in the main text, due to instrument resolution and environmental variation in the Slab2 model, there is a certain amount of uncertainty when measuring the depth of each point on the surface. Specifically, the vertical location of each data point is modeled as a normal distribution in the Slab2 model, where the mean is the preferred depth $d$ and the standard deviation $u$ represents the depth uncertainty. We can further utilize this uncertainty in depth to account for the measurement error and augment our dataset for the correlation analysis. \\

\begin{figure}[t!]
    \centering
    \includegraphics[width=\textwidth]{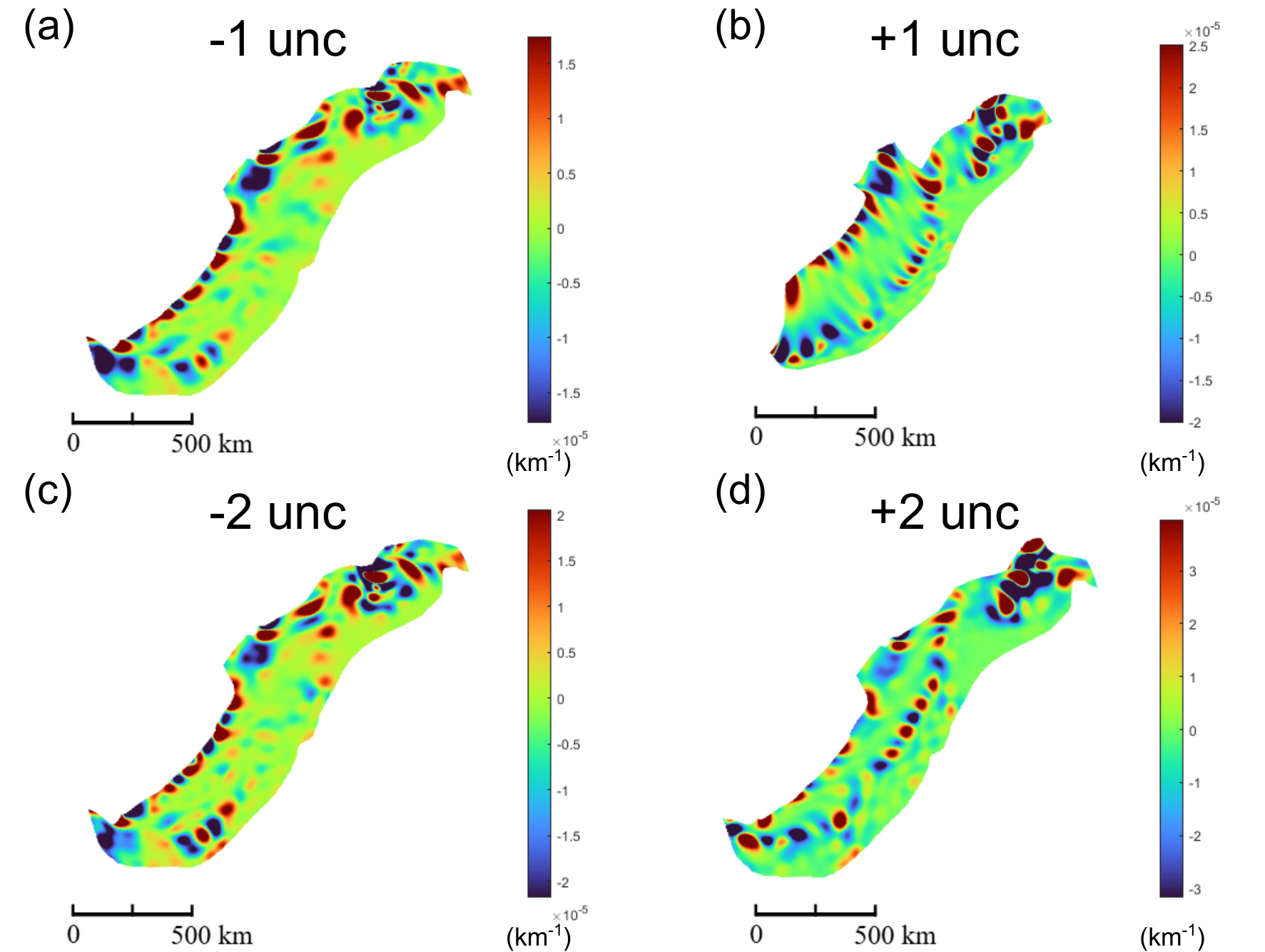}
    \caption{\textbf{The Ryukyu (ryu) surface reconstructed from depth data with $\pm1$ or $\pm2$ uncertainty.} Each surface is color-coded with the Gaussian curvature. (a) $-1$ unc. (b) $+1$ unc. (c) $-2$ unc. (d) $+2$ unc.}
    \label{fig:SI_pm12_unc}
\end{figure}

Here, we consider different levels of uncertainty added to or subtracted from the original depth data and use them for reconstructing the geometries. Specifically, we calculate the mean and Gaussian curvatures using the reconstructed geometries with the modified depth $\tilde{d} = d + k u$, where $d$ is the original depth, $u$ is the given uncertainty, and $k$ is a scalar factor. For instance, Fig.~\ref{fig:SI_pm12_unc} shows the reconstructed surfaces of the Ryukyu (ryu) slab with $-u$, $+u$, $-2u$, and $+2u$ uncertainty values added to the depth respectively. It can be observed that the curvature values of the new surfaces change significantly, while the overall surface geometries still look natural. This approach can therefore provide us with an augmented dataset for our analysis, where we can compute the correlation coefficients between the surface curvature based on the modified depth and the maximum magnitude, average magnitude, and number of earthquakes. \\

\begin{figure*}[t]
    \centering
    \includegraphics[width=\textwidth]{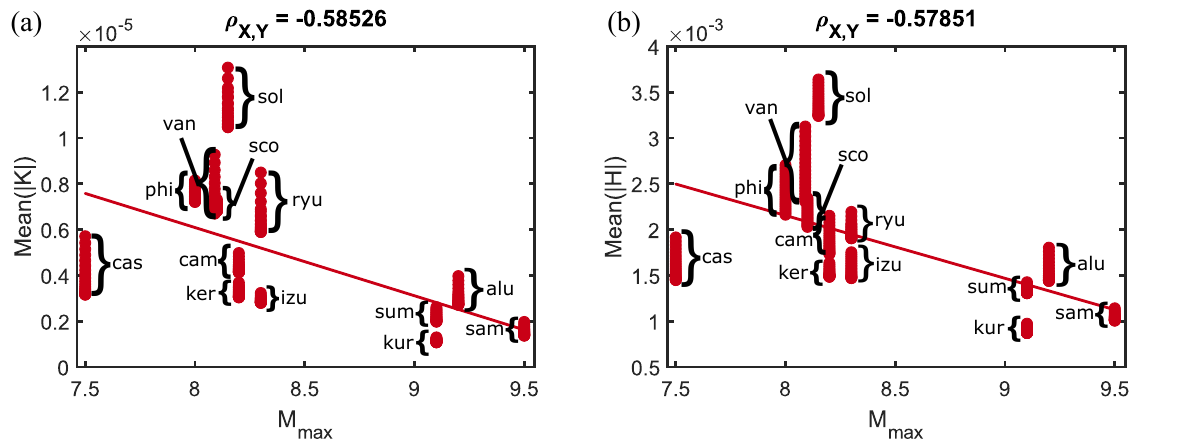}
    \caption{\textbf{Correlation between surface curvature and maximum earthquake magnitude with up to $\pm 1$ uncertainty value added to the depth in the surface reconstruction (273 data points).} (a)~The average absolute Gaussian curvature mean($|K|$). (b)~The average absolute mean curvature mean($|H|$). Each point represents one surface, and the red line is the best-fit straight line. See the caption of main text Fig.~1 for the full name of each model.}
    \label{fig:overall_unc1}
\end{figure*}

\begin{figure*}[t]
    \centering
    \includegraphics[width=\textwidth]{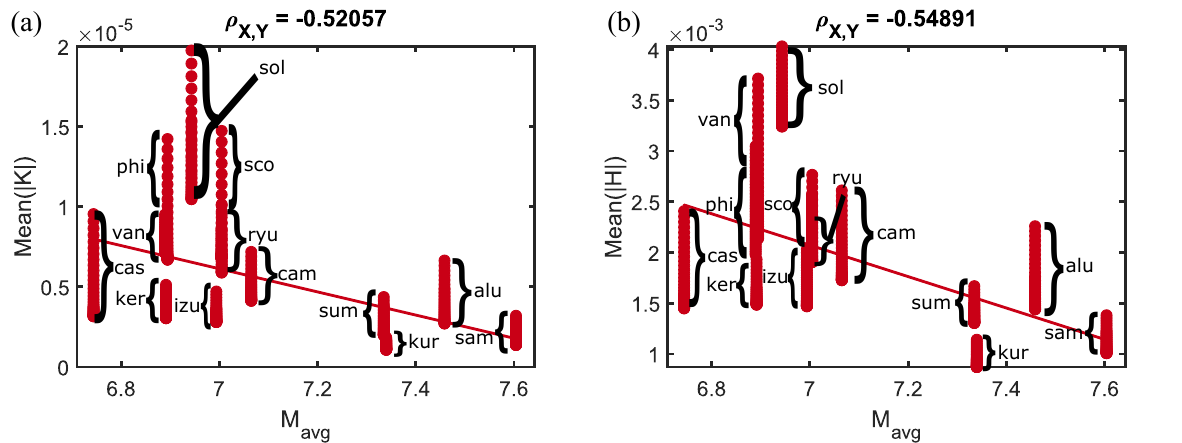}
    \caption{\textbf{Correlation between surface curvature and maximum earthquake magnitude with up to $\pm 2$ uncertainty value added to the depth in the surface reconstruction (533 data points).} (a)~The average absolute Gaussian curvature mean($|K|$). (b)~The average absolute mean curvature mean($|H|$). Each point represents one surface, and the red line is the best-fit straight line. See the caption of main text Fig.~1(b) for the full name of each model.}
    \label{fig:overall_unc2}
\end{figure*}

In Table~\ref{table:uncertainty1.1}--\ref{table:uncertainty2.1}, we consider adding at most $\pm 1$ uncertainty value to the depth by setting different values of $k$ from $-1$ to $1$ with an increment of $0.1$ (13 models $\times$ 21 depth setups, 273 data points in total) and report the detailed correlation coefficients for the mean and Gaussian curvatures with for the maximum magnitude $M_{\text{max}}$, the average magnitude $M_{\text{avg}}$ (with threshold $M_w \geq 6.0$), and the number of earthquakes $n$ (with threshold $M_w \geq 6.0$). In Table~\ref{table:uncertainty1.2}--\ref{table:uncertainty2.2}, we further consider a wider range of $k$ from $-2$ to $2$ with an increment of $0.1$ (13 models $\times$ 41 depth setups, 533 data points in total) and compute the correlation coefficients. From the correlation results, a consistent trend with our previous analysis can be observed, with the average absolute mean and Gaussian curvatures and the maximum and average magnitude of earthquakes still highly negatively correlated even after taking the depth uncertainty into consideration. Also, Fig.~\ref{fig:overall_unc1} and~\ref{fig:overall_unc2} show that in both cases of adding $\pm 1$ or $\pm 2$ uncertainty value to the depth, mean($|K|$) and mean($|H|$) maintain their negative correlation with the maximum magnitude of the earthquakes. This suggests that our findings of the correlation between surface curvature and earthquake productivity are robust to the uncertainty in the depth measurement. \\

\begin{figure}[t!]
    \centering
    \includegraphics[width=\textwidth]{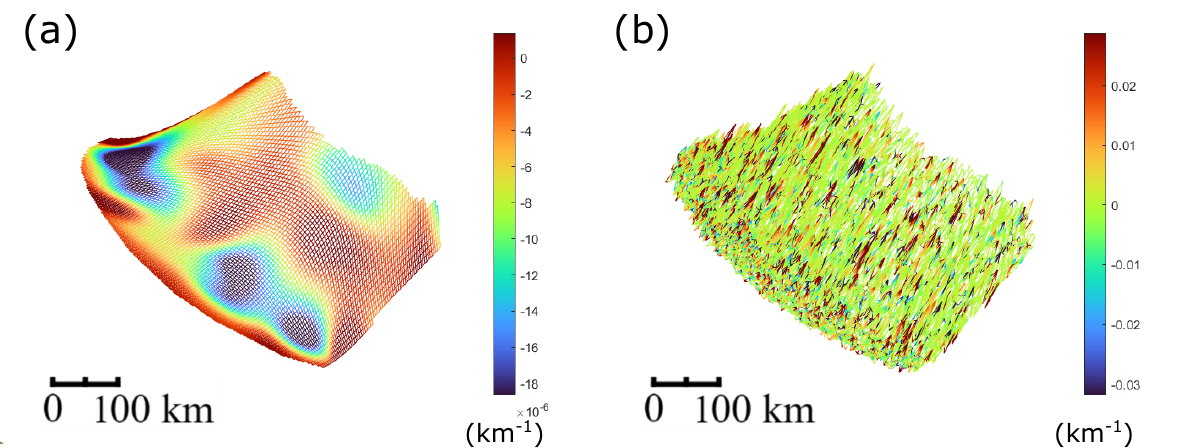}
    \caption{\textbf{The Scotia (sco) surface reconstructed from depth data with random uncertainty.} The color map shows the Gaussian curvature of the surface. (a)~The original slab surface. (b)~The surface reconstructed with random uncertainty.}
    \label{fig:rand_unc}
\end{figure}

Besides adding or subtracting a fixed level of uncertainty, it is natural to ask whether one could also directly add a random uncertainty to the depth data in the surface reconstruction process. However, as shown in Fig.~\ref{fig:rand_unc}, this approach would lead to a highly noisy surface with sharp changes in the surface geometry, and hence the surface curvature extracted from such a surface will be unreliable. Therefore, this approach was not used in our analysis.

\section{Analysis of different characteristic scales}
As described in the main text, we can consider different space separations in the construction of the slab geometry and analyze the surface curvature of the coarsened representation. \\

\begin{figure*}[t!]
    \centering
    \includegraphics[width=\textwidth]{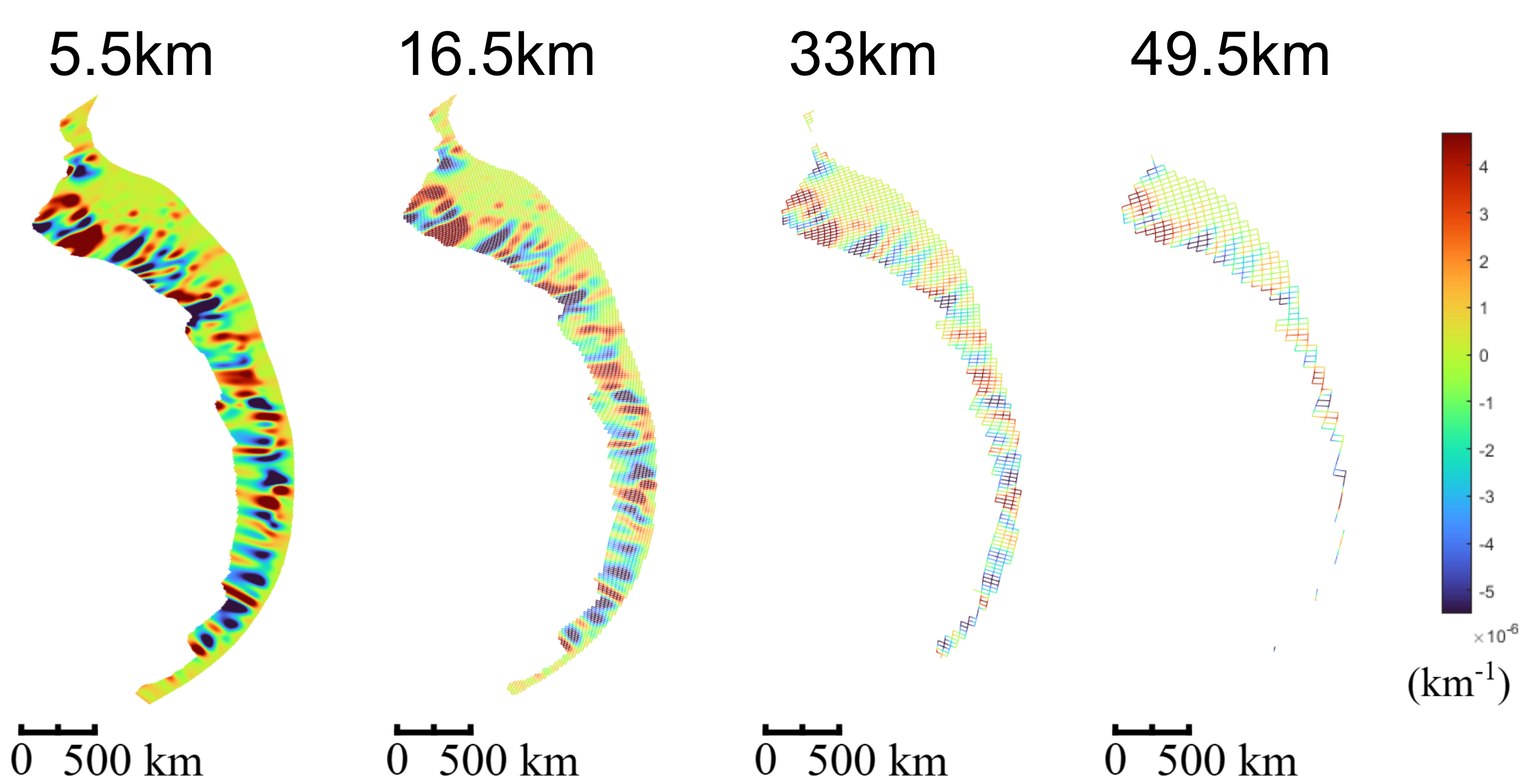}
    \caption{\textbf{Visualization of the Aleutian slab (alu) with different space separations: 5.5km, 16.5km, 33km, and 49.5km.}}
    \label{fig:surface_plotting_space_sep}
\end{figure*}

\begin{figure*}[t!]
    \centering
    \includegraphics[width=\textwidth]{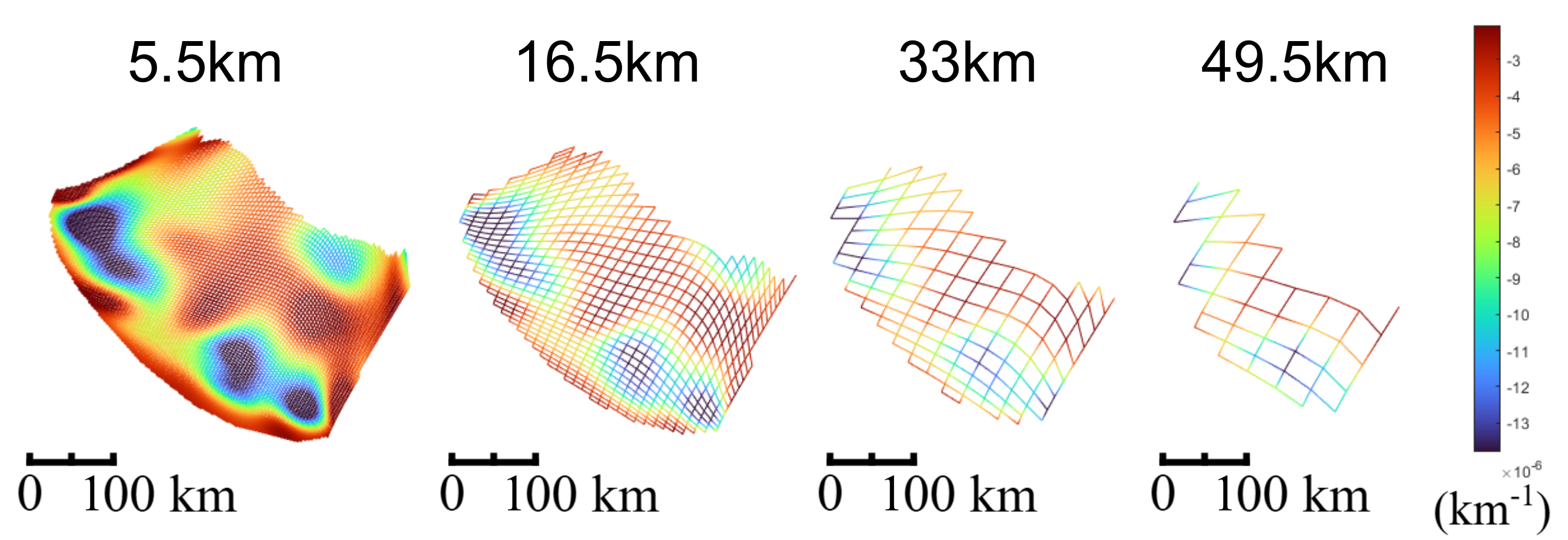}
    \caption{\textbf{Visualization of the Scotia slab (sco) with different space separations: 5.5km, 16.5km, 33km, and 49.5km.}}
    \label{fig:SI_space_sep_sco}
\end{figure*}

\begin{figure*}[t!]
    \centering
    \includegraphics[width=\textwidth]{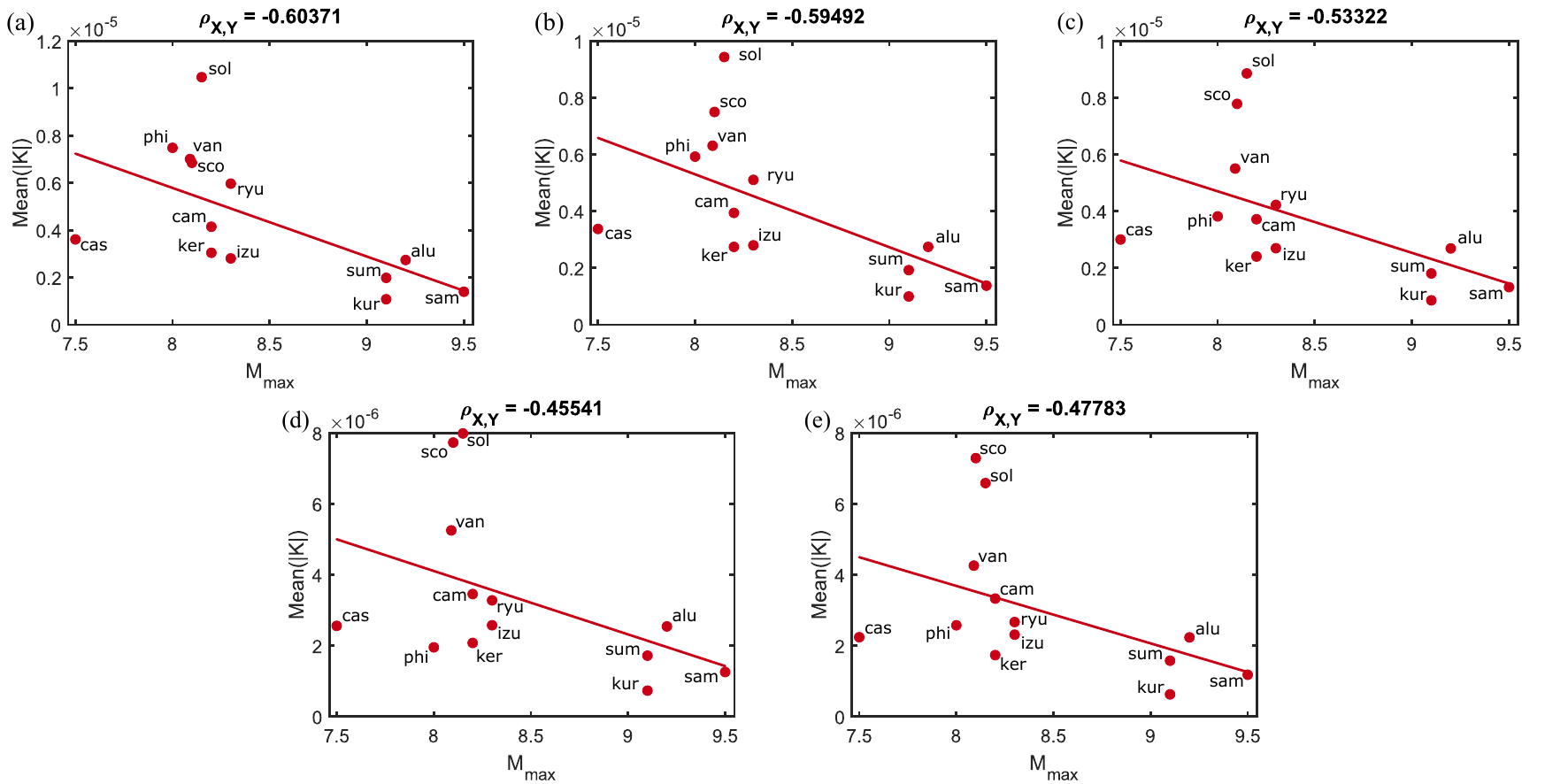}
    \caption{\textbf{Correlation between the average absolute Gaussian curvature $\text{Mean}(|K|)$ and the maximum historically observed magnitude of earthquakes $M_{\max}$ for different space separations.} (a)~5.5km. (b)~16.5km. (c)~27.5km. (d)~38.5km. (e)~49.5km. Each point represents one slab model, and the red line is the best-fit straight line. See the caption of main text Fig.~1 for the full name of each model.}
    \label{fig:curvature_vs_number_space_sep}
\end{figure*}

In Fig.~\ref{fig:surface_plotting_space_sep} and Fig.~\ref{fig:SI_space_sep_sco}, we show two examples (alu and sco) with different space separations. We then repeat the correlation analysis for each choice of the space separation. The detailed results are provided in Table~\ref{table:separation1}--\ref{table:separation3}, from which we can see that the negative correlations between the average absolute curvatures mean($|H|$), mean($|K|$) and both the maximum magnitude $M_{\text{max}}$ and the average magnitude $M_{\text{avg}}$ remain almost unchanged at different characteristic scales. In Fig.~\ref{fig:curvature_vs_number_space_sep}, we can also see that the average absolute Gaussian curvature across different space separations remains negatively correlated with the maximum historically observed magnitude $M_{\max}$. Note that the Solomon slab (sol) appears as an outlier in all plots, which can again be explained by the fact that the geometry of the Solomon slab does not represent the tectonic dynamics as discussed previously. It can also be observed that the Scotia slab (sco) eventually becomes another outlier as the space separation increases. This reflects a tradeoff between the characteristic scaling of the curvature and the scale of the subduction segment itself. The Scotia slab is relatively small, hence the simplified slab surface (see Fig.~\ref{fig:SI_space_sep_sco}) becomes very coarse for larger space separations, making the curvature quantification less meaningful. \\

Besides, as shown in Table~\ref{table:separation3}, the correlation between $\text{CV}(|K|)$ (under the 5.5km separation) and the number of earthquakes $n$ drops rapidly as we increase the space separation. This matches our understanding that the frequency of smaller magnitude events is associated with the roughness of the slab geometries, which is smoothed out as the space separation increases.

\clearpage

\clearpage

\begin{table}[t!]
    \centering
    \begin{tabularx}{\textwidth}{l|Y|Y|Y}
        \toprule
        \diagbox[width=2cm]{$X$}{$Y$}& \textbf{Maximum Magnitude $M_{\max}$} & \textbf{Average Magnitude $M_{\avg}$} & \textbf{Number of Earthquakes $n$}\\
        \hline
        \hline
        Min($H$)    & 0.2314 & 0.2835 & -0.4608 \\
        Min($K$)    & 0.2605 & 0.2801 & -0.0277 \\
        Max($H$)    & -0.0714 & -0.1288 & 0.1988 \\
        Max($K$)    & -0.3194 & -0.3625 & 0.2580 \\
        Mean($H$)   & 0.4536 & 0.4515 & 0.1213 \\
        Mean($K$)   & 0.0488 & -0.0234 & 0.3822 \\
        SD($H$)&	\textbf{-0.6432}&	\textbf{-0.6516}&	-0.2455\\
        SD($K$)&	\textbf{-0.5885}&	\textbf{-0.6144}&	-0.1369\\
        CV($H$)&	-0.0681&	-0.0318&	-0.0099\\
        CV($K$)&	-0.2048&	-0.2540&	-0.0394\\
        \hline
        Min($|H|$)  & -0.2754 & -0.2149 & \textbf{-0.5260} \\
        Min($|K|$)  & -0.2837 & -0.1874 & \textbf{-0.5140} \\
        Max($|H|$)  & -0.1525 & -0.2181 & 0.4771 \\
        Max($|K|$)  & -0.2913 & -0.3401 & 0.1970 \\
        Mean($|H|$) & \textbf{-0.5911} & \textbf{-0.5937} & -0.2226 \\ 
        Mean($|K|$) & \textbf{-0.6037} & \textbf{-0.5979} & -0.2758 \\
        SD($|H|$)&	\textbf{-0.6444}&	\textbf{-0.6581}&	-0.2026\\
        SD($|K|$)&	\textbf{-0.6023}&	\textbf{-0.6274}&	-0.1462\\
        CV($|H|$)&	0.0322&	-0.0187&	0.2908\\
        CV($|K|$)&	0.0773&	-0.0193&	\textbf{0.5276}\\
        \bottomrule
    \end{tabularx}
    \caption{\textbf{A complete table of the correlation coefficients $\rho_{X,Y}$ between the surface curvature quantities ($X$) and earthquake productivity measurements ($Y$) of subduction zones.} For each geometric quantity, we calculate the Pearson Correlation coefficients between it and the maximum historically observed magnitude $M_{\max}$, the average magnitude $M_{\avg}$, and the number of earthquakes $n$ for 13 slab geometry models from the Slab2 dataset with magnitude threshold $M_w \geq 6.0$. Correlation coefficients with absolute value $\geq 0.5$ are in boldface.}
    \label{tab:SI_full_statistics}
\end{table}

\begin{table*}[t!]
    \centering
    \begin{tabularx}{\textwidth}{l|YYYYY}
    \toprule
    \multirow{2}{*}{\diagbox[width=2cm]{$X$}{$Y$}} & \multicolumn{5}{c}{\textbf{Average magnitude $M_{\text{avg}}$ with different thresholds}}\\
    \cmidrule(lr){2-6}
     & $M_w \geq 5.0$ & $M_w \geq  5.5$ & $M_w \geq 6.0$ & $M_w \geq 6.5$ & $M_w \geq 7.0$ \\ \midrule
     \midrule
    Min($H$) & 0.1565&	0.2077&	0.2835&	0.2734&	0.2800\\
    Min($K$) & 0.1518&	0.2011&	0.2801&	0.2966&	0.2528\\
    Max($H$) & -0.1153&	-0.1176&-0.1288&-0.0983&-0.0189\\
    Max($K$) & -0.2377&	-0.2801&-0.3625&-0.3717&-0.3540\\
    Mean($H$) & 0.3684&	0.4009&	0.4515&	0.4847&	0.4840\\
    Mean($K$) & 0.1899&	0.1192&	-0.0234&-0.0908&-0.1568\\
    SD($H$)&	\textbf{-0.6323}&	\textbf{-0.6370}&	\textbf{-0.6516}&	\textbf{-0.6440}&	-0.3170\\
    SD($K$)&	\textbf{-0.5119}&	\textbf{-0.5517}&	\textbf{-0.6144}&	\textbf{-0.6329}&	-0.1883\\
    CV($H$)&	0.0350&	0.0020&	-0.0318&	-0.0715&	0.0713\\
    CV($K$)&	-0.3142&	-0.2606&	-0.2540&	-0.2545&	-0.0857\\
    \midrule
    Min($|H|$)  & -0.4124&	-0.3468&	-0.2149&	-0.1694&	-0.1177\\
    Min($|K|$) & -0.3592&	-0.3125&	-0.1874&	-0.1371&	-0.0866\\
    Max($|H|$)  & -0.0962&	-0.1477&	-0.2181&	-0.1996&	-0.1830\\
    Max($|K|$)  & -0.2192&	-0.2616&	-0.3401&	-0.3463&	-0.3103\\
    Mean($|H|$) & \textbf{-0.5576}&	\textbf{-0.5692}&	\textbf{-0.5937}&	\textbf{-0.6042}&	\textbf{-0.5630}\\
    Mean($|K|$) & \textbf{-0.5720}&	\textbf{-0.5833}&	\textbf{-0.5979}&	\textbf{-0.5989}&	\textbf{-0.5473}\\
    SD($|H|$)&	\textbf{-0.5802}&	\textbf{-0.6119}&	\textbf{-0.6581}&	\textbf{-0.6676}&	-0.2683\\
    SD($|K|$)&	\textbf{-0.5303}&	\textbf{-0.5694}&	\textbf{-0.6274}&	\textbf{-0.6411}&	-0.2066\\
    CV($|H|$)&	0.1083&	0.0470&	-0.0187&	-0.0008&	0.1787\\
    CV($|K|$)&	0.1018&	0.0552&	-0.0193&	-0.0172&	0.4436\\
    \bottomrule
    \end{tabularx}
    \caption{\textbf{Correlation coefficients $\rho_{X,Y}$ between the mean and Gaussian curvatures and the average earthquake magnitude $M_{\text{avg}}$ with different magnitude thresholds $M_w \geq 5.0, 5.5, 6.0, 6.5, 7.0$.}  Coefficients with absolute value greater than 0.5 are in boldface.} 
    \label{tab:SI_avgmag}
\end{table*}

\begin{table*}[t!]
    \centering
    \begin{tabularx}{\textwidth}{l|YYYYY}
    \toprule
    \multirow{2}{*}{\diagbox[width=2cm]{$X$}{$Y$}} & \multicolumn{5}{c}{\textbf{Number of earthquakes $n$ with different thresholds}}\\
    \cmidrule(lr){2-6}
     & $M_w \geq 5.0$ & $M_w \geq  5.5$ & $M_w \geq 6.0$ & $M_w \geq 6.5$ & $M_w \geq 7.0$ \\ \midrule
     \midrule
    min($H$)&	-0.3048&	-0.3839&	-0.4608&	-0.4099&	-0.3856\\
    min($K$)&	0.0606&	0.0375&	-0.0277&	-0.0389&	0.0376\\
    max($H$)&	0.2298&	0.2214&	0.1988&	0.1554&	0.0260\\
    max($K$)&	0.1733&	0.1893&	0.2580&	0.2429&	0.1790\\
    mean($H$)&	0.1860&	0.1961&	0.1213&	0.0873&	0.1086\\
    mean($K$)&	0.2929&	0.3223&	0.3822&	0.4169&	0.4096\\
    SD($H$)&	-0.2321&	-0.2805&	-0.2455&	-0.2572&	-0.3170\\
    SD($K$)&	-0.1754&	-0.2004&	-0.1369&	-0.1348&	-0.1883\\
    CV($H$)&	-0.1029&	-0.0657&	-0.0099&	0.0344&	0.0713\\
    CV($K$)&	0.0934&	-0.0190&	-0.0394&	-0.0227&	-0.0857\\
    \midrule
    min($|H|$)&	-0.3747&	-0.4662&	\textbf{-0.5260}&	\textbf{-0.5522}&	\textbf{-0.5337}\\
    min($|K|$)&	-0.4272&	-0.4739&	\textbf{-0.5140}&	\textbf{-0.5422}&	\textbf{-0.5142}\\
    max($|H|$)&	0.3361&	0.4143&	0.4771&	0.4163&	0.3556\\
    max($|K|$)&	0.1196&	0.1338&	0.1970&	0.1816&	0.1020\\
    mean($|H|$)&	-0.2410&	-0.2782&	-0.2226&	-0.2075&	-0.2525\\
    mean($|K|$)&	-0.2956&	-0.3314&	-0.2758&	-0.2722&	-0.3178\\
    SD($|H|$)&	-0.2226&	-0.2565&	-0.2026&	-0.2102&	-0.2683\\
    SD($|K|$)&	-0.1749&	-0.2029&	-0.1462&	-0.1518&	-0.2066\\
    CV($|H|$)&	0.2505&	0.3019&	0.2908&	0.1988&	0.1787\\
    CV($|K|$)&	\textbf{0.5384}&	\textbf{0.5588}&	\textbf{0.5276}&	0.4611&	0.4436\\
    \bottomrule
    \end{tabularx}
    \caption{\textbf{Correlation coefficients $\rho_{X,Y}$  between the mean and Gaussian curvatures and the number of earthquakes $n$ with different magnitude thresholds $M_w \geq 5.0, 5.5, 6.0, 6.5, 7.0$.}  Coefficients with absolute value greater than 0.5 are in boldface.}
    \label{tab:SI_num}
\end{table*}

\begin{table*}[t!]
    \centering
    \begin{tabularx}{\textwidth}{l|YYYY}
    \toprule
    \multirow{2}{*}{\diagbox[width=2cm]{$X$}{$Y$}} & \multicolumn{4}{c}{\textbf{Maximum magnitude $M_{\text{max}}$ (with uncertainty added)}}\\
    \cmidrule(lr){2-5}
    & -1 unc & original & +1 unc & Overall\\ \midrule
    \midrule
    Min($H$)&	0.2045&	0.2314&	0.2785&	0.2540\\
    Min($K$)&	0.2100&	0.2605&	0.4240&	0.2728\\
    Max($H$)&	0.0125&	-0.0714&	-0.1578&	-0.0787\\
    Max($K$)&	-0.2170&	-0.3194&	-0.3820&	-0.3089\\
    Mean($H$)&	0.4181&	0.4536&	0.4775&	0.4446\\
    Mean($K$)&	0.0834&	0.0488&	-0.0052&	0.0449\\
    SD($H$)&	-0.4854&	\textbf{-0.6432}&	\textbf{-0.7144}&	\textbf{-0.6221}\\
    SD($K$)&	-0.4600&	\textbf{-0.5885}&	\textbf{-0.6127}&	\textbf{-0.5660}\\
    CV($H$)&	-0.0704&	-0.0681&	-0.0810&	-0.0685\\
    CV($K$)&	-0.1599&	-0.2048&	-0.2201&	-0.0671\\
    \midrule
    Min($|H|$)&	-0.3618&	-0.2754&	-0.2411&	-0.3215\\
    Min($|K|$)&	-0.4257&	-0.2837&	-0.3398&	-0.2430\\
    Max($|H|$)&	-0.1576&	-0.1525&	-0.3397&	-0.2042\\
    Max($|K|$)&	-0.2268&	-0.2913&	-0.3910&	-0.2976\\
    Mean($|H|$)&	\textbf{-0.5227}&	\textbf{-0.5911}&	\textbf{-0.6223}&	\textbf{-0.5785}\\
    Mean($|K|$)&	\textbf{-0.5155}&	\textbf{-0.6037}&	\textbf{-0.6149}&	\textbf{-0.5853}\\
    SD($|H|$)&	-0.4965&	\textbf{-0.6444}&	\textbf{-0.7284}&	\textbf{-0.6266}\\
    SD($|K|$)&	-0.4597&	\textbf{-0.6023}&	\textbf{-0.6214}&	\textbf{-0.5766}\\
    CV($|H|$)&	0.0754&	0.0322&	-0.0725&	0.0073\\
    CV($|K|$)&	0.2098&	0.0773&	-0.0927&	0.0632\\
    \bottomrule
    \end{tabularx}
    \caption{\textbf{Correlation coefficients $\rho_{X,Y}$ between the mean and Gaussian curvatures of the surfaces generated with the uncertainty added and the maximum magnitude of earthquakes $M_{\max}$.}  Coefficients with absolute value greater than 0.5 are in boldface.}
    \label{table:uncertainty1.1}
\end{table*}

\begin{table*}[t!]
    \centering
    \begin{tabularx}{\textwidth}{l|YYYY}
    \toprule
    \multirow{2}{*}{\diagbox[width=2cm]{$X$}{$Y$}} & \multicolumn{4}{c}{\textbf{Average magnitude $M_{\text{avg}}$ (with uncertainty added)}}\\
    \cmidrule(lr){2-5}
    & -1 unc & original & +1 unc & Overall\\ \midrule
    \midrule
    Min($H$)&		0.2601&	0.2835&	0.2730&		0.2943\\
    Min($K$)&		0.2491&	0.2801&	0.4361&		0.2940\\
    Max($H$)&		-0.0839&	-0.1288&	-0.1717& -0.1382\\
    Max($K$)&		-0.2955&	-0.3625&	-0.4310&	      	-0.3619\\
    Mean($H$)&		0.4158&	0.4515&	0.4755&		0.4426\\
    Mean($K$)&		0.0109&	-0.0234&	-0.0777&		-0.0273\\
    SD($H$)&	\textbf{-0.5038}&	\textbf{-0.6516}&	\textbf{-0.7100}&	\textbf{-0.6299}\\
    SD($K$)&	\textbf{-0.5032}&	\textbf{-0.6144}&	\textbf{-0.6287}&	\textbf{-0.5924}\\
    CV($H$)&	-0.0356&	-0.0318&	-0.0506&	-0.0349\\
    CV($K$)&	-0.1448&	-0.2540&	-0.1461&	-0.0418\\

    \midrule
    Min($|H|$)&		-0.3665&	-0.2149&	-0.1474&		-0.3026\\
    Min($|K|$)&		-0.3650&	-0.1874&	-0.3405&		-0.2019\\
    Max($|H|$)&		-0.2419&	-0.2181&	-0.3688&		-0.2693\\
    Max($|K|$)&		-0.2988&	-0.3401&	-0.4345&		-0.3506\\
    Mean($|H|$)&		\textbf{-0.5294}&	\textbf{-0.5937}&	\textbf{-0.6197}&		\textbf{-0.5807}\\
    Mean($|K|$)&		\textbf{-0.5204}&	\textbf{-0.5979}&	\textbf{-0.6097}&		\textbf{-0.5815}\\
    SD($|H|$)&	\textbf{-0.5218}&	\textbf{-0.6581}&	\textbf{-0.7318}&	\textbf{-0.6402}\\
    SD($|K|$)&	\textbf{-0.5046}&	\textbf{-0.6274}&	\textbf{-0.6373}&	\textbf{-0.6028}\\
    CV($|H|$)&	-0.0015&	-0.0187&	-0.1068&	-0.0431\\
    CV($|K|$)&	0.1041&	-0.0193&	-0.1700&	-0.0305\\
    \bottomrule
    \end{tabularx}
    \caption{\textbf{Correlation coefficients $\rho_{X,Y}$ between the mean and Gaussian curvatures of the surfaces generated with the uncertainty added and the average magnitude of earthquakes $M_{\avg}$.} Here, the average magnitude data is obtained based on the magnitude threshold $M_w \geq 6.0$. Coefficients with absolute value greater than 0.5 are in boldface.}
    \label{table:uncertainty1.5.1}
\end{table*}

\begin{table*}[t]
    \centering
    \begin{tabularx}{\textwidth}{l|YYYY}
    \toprule
    \multirow{2}{*}{\diagbox[width=2cm]{$X$}{$Y$}} & \multicolumn{4}{c}{\textbf{Number of earthquakes $n$ (with uncertainty added)}}\\
    \cmidrule(lr){2-5}
     & -1 unc & original & +1 unc & Overall \\ \midrule
     \midrule
    Min($H$)&	-0.4880&	-0.4608&	-0.1217&	-0.3889\\
    Min($K$)&	-0.1918&	-0.0277&	0.0734&	-0.0350\\
    Max($H$)&	0.4214&	0.1988&	-0.0170&	0.2109\\
    Max($K$)&	0.4369&	0.2580&	0.1767&	0.2820\\
    Mean($H$)&	0.0797&	0.1213&	0.1563&	0.1197\\
    Mean($K$)&	0.3883&	0.3822&	0.3628&	0.3794\\
    SD($H$)&	-0.1320&	-0.2455&	-0.3237&	-0.2374\\
    SD($K$)&	0.0261&	-0.1369&	-0.1681&	-0.1099\\
    CV($H$)&	0.0255&	-0.0099&	-0.0257&	-0.0031\\
    CV($K$)&	-0.0541&	-0.0394&	-0.1742&	-0.0616\\
    \midrule
    Min($|H|$)&	-0.2007&	\textbf{-0.5260}&	-0.1508&	-0.2406\\
    Min($|K|$)&	\textbf{-0.5450}&	\textbf{-0.5140}&	-0.1660&	-0.2998\\
    Max($|H|$)&	\textbf{0.5368}&	0.4771&	0.2429&	0.4391\\
    Max($|K|$)&	0.3962&	0.1970&	0.1486&	0.2287\\
    Mean($|H|$)&	-0.1782&	-0.2226&	-0.2695&	-0.2223\\
    Mean($|K|$)&	-0.1733&	-0.2758&	-0.2749&	-0.2531\\
    SD($|H|$)&	-0.0814&	-0.2026&	-0.2840&	-0.1932\\
    SD($|K|$)&	0.0390&	-0.1462&	-0.1641&	-0.1114\\
    CV($|H|$)&	0.4128&	0.2908&	0.1860&	0.2804\\
    CV($|K|$)&	\textbf{0.6839}&	\textbf{0.5276}&	0.3709&	\textbf{0.5305}\\
    \bottomrule
    \end{tabularx}
    \caption{\textbf{Correlation coefficients $\rho_{X,Y}$ between the mean and Gaussian curvatures of the surfaces generated with the uncertainty added and the number of earthquakes $n$.} Here, the earthquake numbers are obtained based on the magnitude threshold $M_w \geq 6.0$. Coefficients with absolute value greater than 0.5 are in boldface.}
    \label{table:uncertainty2.1}
\end{table*}

\begin{table*}[t]
    \centering
    \begin{tabularx}{\textwidth}{l|YYYYYY}
    \toprule
    \multirow{2}{*}{\diagbox[width=2cm]{$X$}{$Y$}} & \multicolumn{6}{c}{\textbf{Maximum magnitude $M_{\text{max}}$ (with uncertainty added)}}\\
    \cmidrule(lr){2-7}
    & -2 unc & -1 unc & original & +1 unc & +2 unc & Overall\\ \midrule
    \midrule
    Min($H$)&	0.1331&	0.2045&	0.2314&	0.2785&	0.1414&	0.1831\\
    Min($K$)&	0.0500&	0.2100&	0.2605&	0.4240&	0.3424&	0.2362\\
    Max($H$)&	0.1599&	0.0125&	-0.0714&	-0.1578&	-0.1863&	-0.0222\\
    Max($K$)&	-0.0612&	-0.2170&	-0.3194&	-0.3820&	-0.1784&	-0.1773\\
    Mean($H$)&	0.3679&	0.4181&	0.4536&	0.4775&	0.4945&	0.4233\\
    Mean($K$)&	0.0970&	0.0834&	0.0488&	-0.0052&	-0.0746&	0.0347\\
    SD($H$)&	-0.3347&	-0.4854&	\textbf{-0.6432}&	\textbf{-0.7144}&	\textbf{-0.6875}&	\textbf{-0.5572}\\
    SD($K$)&	-0.2492&	-0.4600&	\textbf{-0.5885}&	\textbf{-0.6127}&	\textbf{-0.5099}&	-0.4508\\
    CV($H$)&	-0.0211&	-0.0704&	-0.0681&	-0.0810&	-0.1246&	-0.0541\\
    CV($K$)&	-0.0850&	-0.1599&	-0.2048&	-0.2201&	-0.1128&	-0.0686\\
    \midrule
    Min($|H|$)&	-0.4560&	-0.3618&	-0.2754&	-0.2411&	-0.2582&	-0.3232\\
    Min($|K|$)&	-0.3542&	-0.4257&	-0.2837&	-0.3398&	-0.4354&	-0.2673\\
    Max($|H|$)&	0.1075&	-0.1576&	-0.1525&	-0.3397&	-0.1791&	-0.1159\\
    Max($|K|$)&	-0.0679&	-0.2268&	-0.2913&	-0.3910&	-0.1755&	-0.1803\\
    Mean($|H|$)&	-0.4263&	\textbf{-0.5227}&	\textbf{-0.5911}&	\textbf{-0.6223}&	\textbf{-0.6422}&	\textbf{-0.5480}\\
    Mean($|K|$)&	-0.3951&	\textbf{-0.5155}&	\textbf{-0.6037}&	\textbf{-0.6149}&	\textbf{-0.5621}& \textbf{-0.5203}\\
    SD($|H|$)&	-0.3360&	-0.4965&	\textbf{-0.6444}&	\textbf{-0.7284}&	\textbf{-0.7224}&	\textbf{-0.5642}\\
    SD($|K|$)&	-0.2271&	-0.4597&	\textbf{-0.6023}&	\textbf{-0.6214}&	-0.4931&	-0.4394\\
    CV($|H|$)&	0.1753&	0.0754&	0.0322&	-0.0725&	-0.0865&	0.0095\\
    CV($|K|$)&	0.2977&	0.2098&	0.0773&	-0.0927&	0.0294&	0.1077\\
    \bottomrule
    \end{tabularx}
    \caption{\textbf{Correlation coefficients $\rho_{X,Y}$ between the mean and Gaussian curvatures of the surfaces generated with the uncertainty added and the maximum magnitude of earthquakes $M_{\max}$.} Coefficients with absolute value greater than 0.5 are in boldface.}
    \label{table:uncertainty1.2}
\end{table*}

\begin{table*}[t]
    \centering
    \begin{tabularx}{\textwidth}{l|YYYYYY}
    \toprule
    \multirow{2}{*}{\diagbox[width=2cm]{$X$}{$Y$}} & \multicolumn{6}{c}{\textbf{Average magnitude $M_{\text{avg}}$ (with uncertainty added)}}\\
    \cmidrule(lr){2-7}
    & -2 unc & -1 unc & original & +1 unc & +2 unc & Overall\\ \midrule
    \midrule
    Min($H$)&	0.1917&	0.2601&	0.2835&	0.2730&	0.1175&	0.2065\\
    Min($K$)&	0.1077&	0.2491&	0.2801&	0.4361&	0.3678&	0.2628\\
    Max($H$)&	0.0802&	-0.0839&	-0.1288&	-0.1717&	-0.1775&	-0.0715\\
    Max($K$)&	-0.1484&	-0.2955&	-0.3625&	-0.4310&	-0.2217&	-0.2302\\
    Mean($H$)&	0.3659&	0.4158&	0.4515&	0.4755&	0.4916&	-0.2302\\
    Mean($K$)&	0.0235&	0.0109&	-0.0234&	-0.0777&	-0.1471&	-0.0375\\
    SD($H$)&	-0.3561&	\textbf{-0.5038}&	\textbf{-0.6516}&	\textbf{-0.7100}&	\textbf{-0.6743}&	\textbf{-0.5631}\\
    SD($K$)&	-0.3043&	\textbf{-0.5032}&	\textbf{-0.6144}&	\textbf{-0.6287}&	\textbf{-0.5262}&	-0.4779\\
    CV($H$)&	0.0080&	-0.0356&	-0.0318&	-0.0506&	-0.1029&	-0.0277\\
    CV($K$)&	-0.0878&	-0.1448&	-0.2540&	-0.1461&	-0.0537&	-0.0394\\
    \midrule
    Min($|H|$)&	-0.4123&	-0.3665&	-0.2149&	-0.1474&	-0.2638&	-0.3080\\
    Min($|K|$)&	-0.3572&	-0.3650&	-0.1874&	-0.3405&	-0.4290&	-0.2303\\
    Max($|H|$)&	0.0208&	-0.2419&	-0.2181&	-0.3688&	-0.1812&	-0.1658\\
    Max($|K|$)&	-0.1540&	-0.2988&	-0.3401&	-0.4345&	-0.2191&	-0.2324\\
    Mean($|H|$)&	-0.4309&	\textbf{-0.5294}&	\textbf{-0.5937}&	\textbf{-0.6197}&	\textbf{-0.6358}&	\textbf{-0.5489}\\
    Mean($|K|$)&	-0.4084&	\textbf{-0.5204}&	\textbf{-0.5979}&	\textbf{-0.6097}&	\textbf{-0.5628}&	\textbf{-0.5206}\\
    SD($|H|$)&	-0.3701&	\textbf{-0.5218}&	\textbf{-0.6581}&	\textbf{-0.7318}&	\textbf{-0.7152}&	\textbf{-0.5770}\\
    SD($|K|$)&	-0.2850&	\textbf{-0.5046}&	\textbf{-0.6274}&	\textbf{-0.6373}&	\textbf{-0.5110}&	-0.4672\\
    CV($|H|$)&	0.0710&	-0.0015&	-0.0187&	-0.1068&	-0.1006&	-0.0394\\
    CV($|K|$)&	0.1874&	0.1041&	-0.0193&	-0.1700&	-0.0311&	0.0201\\
    \bottomrule
    \end{tabularx}
    \caption{\textbf{Correlation coefficients $\rho_{X,Y}$ between the mean and Gaussian curvatures of the surfaces generated with the uncertainty added and the average magnitude of earthquakes $M_{\avg}$.} Here, the average magnitude data is obtained based on the magnitude threshold $M_w \geq 6.0$. Coefficients with absolute value greater than 0.5 are in boldface.}
    \label{table:uncertainty1.5.2}
\end{table*}

\begin{table*}[t]
    \centering
    \begin{tabularx}{\textwidth}{l|YYYYYY}
    \toprule
    \multirow{2}{*}{\diagbox[width=2cm]{$X$}{$Y$}} & \multicolumn{6}{c}{\textbf{Number of earthquakes $n$ (with uncertainty added)}}\\
    \cmidrule(lr){2-7}
    & -2 unc & -1 unc & original & +1 unc & +2 unc & Overall \\ \midrule \midrule
    Min($H$)&	\textbf{-0.5577}&	-0.4880&	-0.4608&	-0.1217&	0.0845&	-0.2677\\
    Min($K$)&	-0.2111&	-0.1918&	-0.0277&	0.0734&	0.0153&	-0.0629\\
    Max($H$)&	0.3847&	0.4214&	0.1988&	-0.0170&	0.0247&	0.2071\\
    Max($K$)&	0.2238&	0.4369&	0.2580&	0.1767&	0.0340&	0.1908\\
    Mean($H$)&	0.0283&	0.0797&	0.1213&	0.1563&	0.1890&	0.1162\\
    Mean($K$)&	0.3771&	0.3883&	0.3822&	0.3628&	0.3178&	0.3697\\
    SD($H$)&	-0.0290&	-0.1320&	-0.2455&	-0.3237&	-0.3377&	-0.2109\\
    SD($K$)&	0.0856&	0.0261&	-0.1369&	-0.1681&	-0.1447&	-0.0679\\
    CV($H$)&	0.0943&	0.0255&	-0.0099&	-0.0257&	-0.0468&	0.0168\\
    CV($K$)&	0.2467&	-0.0541&	-0.0394&	-0.1742&	0.0320&	-0.0229\\
    \midrule
    Min($|H|$)&	-0.3967&	-0.2007&	\textbf{-0.5260}&	-0.1508&	0.1633&	-0.2256\\
    Min($|K|$)&	-0.2983&	\textbf{-0.5450}&	\textbf{-0.5140}&	-0.1660&	-0.0409&	-0.3107\\
    Max($|H|$)&	0.4430&	\textbf{0.5368}&	0.4771&	0.2429&	-0.0036&	0.3067\\
    Max($|K|$)&	0.2171&	0.3962&	0.1970&	0.1486&	0.0405&	0.1756\\
    Mean($|H|$)&	-0.1288&	-0.1782&	-0.2226&	-0.2695&	-0.3051&	-0.2177\\
    Mean($|K|$)&	-0.0696&	-0.1733&	-0.2758&	-0.2749&	-0.2259&	-0.2056\\
    SD($|H|$)&	0.0358&	-0.0814&	-0.2026&	-0.2840&	-0.3256&	-0.1696\\
    SD($|K|$)&	0.0955&	0.0390&	-0.1462&	-0.1641&	-0.1284&	-0.0588\\
    CV($|H|$)&	\textbf{0.5530}&	0.4128&	0.2908&	0.1860&	0.0957&	0.2671\\
    CV($|K|$)&	\textbf{0.5765}&	\textbf{0.6839}&	\textbf{0.5276}&	0.3709&	0.3105&	0.4886\\
    \bottomrule
    \end{tabularx}
    \caption{\textbf{Correlation coefficients $\rho_{X,Y}$ between the mean and Gaussian curvatures of the surfaces generated with the uncertainty added and the number of earthquakes $n$.} Here, the earthquake numbers are obtained based on the magnitude threshold $M_w \geq 6.0$. Coefficients with absolute value greater than 0.5 are in boldface.}
    \label{table:uncertainty2.2}
\end{table*}

\begin{table*}[t]
    \centering
    \small
    \setlength{\tabcolsep}{4pt}
     \begin{tabularx}{\textwidth}{l|YYYYY}
    \toprule
    \multirow{2}{*}{\diagbox[width=2cm]{$X$}{$Y$}} & \multicolumn{5}{c}{\textbf{Maximum magnitude $M_{\text{max}}$ (with different space separations)}}\\
    \cmidrule(lr){2-6}
    & 5.5km & 16.5km & 27.5km & 38.5km & 49.5km\\ \midrule \midrule
    Min($H$)&	0.2314&	0.3774&	0.4224&	0.2213&	0.2119\\
    Min($K$)&	0.2605&	0.2593&	0.3461&	0.2251&	0.3350\\
    Max($H$)&	-0.0714&	-0.0166&	0.0010&	0.1148&	0.3002\\
    Max($K$)&	-0.3194&	-0.3407&	-0.3359&	-0.1818&	-0.1173\\
    Mean($H$)&	0.4536&	0.4188&	0.3709&	0.3612&	0.3946\\
    Mean($K$)&	0.0488&	0.1130&	0.1577&	0.1595&	0.1933\\
    SD($H$)&	\textbf{-0.6432}&	\textbf{-0.6245}&	\textbf{-0.5021}&	-0.3242&	-0.2472\\
    SD($K$)&	\textbf{-0.5885}&	\textbf{-0.6300}&	\textbf{-0.6026}&	-0.4513&	-0.4971\\
    CV($H$)&	-0.0681&	-0.1187&	-0.1482&	-0.2154&	-0.2760\\
    CV($K$)&	-0.2048&	-0.3330&	-0.2065&	-0.1795&	-0.2407\\
    \midrule
    Min($|H|$)&	-0.2754&	-0.3933&	\textbf{-0.5770}&	-0.2600&	-0.2548\\
    Min($|K|$)&	-0.2837&	-0.3762&	-0.3038&	-0.3468&	-0.2297\\
    Max($|H|$)&	-0.1525&	-0.2859&	-0.3579&	-0.1723&	-0.1217\\
    Max($|K|$)&	-0.2913&	-0.3436&	-0.4423&	-0.2833&	-0.2288\\
    Mean($|H|$)&	\textbf{-0.5911}&	\textbf{-0.5429}&	-0.4677&	-0.4271&	-0.4266\\
    Mean($|K|$)&	\textbf{-0.6037}&	\textbf{-0.5949}&	\textbf{-0.5332}&	-0.4554&	-0.4778\\
    SD($|H|$)&	\textbf{-0.6444}&	\textbf{-0.6226}&	\textbf{-0.5098}&	-0.3255&	-0.2782\\
    SD($|K|$)&	\textbf{-0.6023}&	\textbf{-0.6669}&	\textbf{-0.6526}&	-0.4830&	\textbf{-0.5452}\\
    CV($|H|$)&	0.0322&	-0.0143&	0.0591&	0.2391&	0.2308\\
    CV($|K|$)&	0.0773&	0.0589&	-0.0891&	0.1619&	0.0825\\
    \bottomrule
    \end{tabularx}
    \caption{\textbf{Correlation coefficients $\rho_{X,Y}$ between the surface curvatures with different space separations and the maximum earthquake magnitude of subduction zones.} Note that 5.5km is the original space separation. Coefficients with absolute value greater than 0.5 are in boldface.}
    \label{table:separation1}
\end{table*}

\begin{table*}[t]
    \centering
    \small
    \setlength{\tabcolsep}{4pt}
     \begin{tabularx}{\textwidth}{l|YYYYY}
    \toprule
    \multirow{2}{*}{\diagbox[width=2cm]{$X$}{$Y$}} & \multicolumn{5}{c}{\textbf{Average magnitude $M_{\text{avg}}$ (with different space separations)}}\\
    \cmidrule(lr){2-6}
    & 5.5km & 16.5km & 27.5km & 38.5km & 49.5km\\ \midrule \midrule
    Min($H$)&	0.2835&	0.4137&	0.4217&	0.1954&	0.1785\\
    Min($K$)&	0.2801&	0.3343&	0.4013&	0.2464&	0.3453\\
    Max($H$)&	-0.1288&	-0.0846&	-0.0709&	0.0338&	0.2399\\
    Max($K$)&	-0.3625&	-0.3986&	-0.3828&	-0.2123&	-0.0821\\
    Mean($H$)&	0.4515&	0.4187&	0.3686&	0.3495&	0.3741\\
    Mean($K$)&	-0.0234&	0.0460&	0.0952&	0.1037&	0.1440\\
    SD($H$)&	\textbf{-0.6516}&	\textbf{-0.6277}&	-0.4991&	-0.3142&	-0.2321\\
    SD($K$)&	\textbf{-0.6144}&	\textbf{-0.6446}&	\textbf{-0.6031}&	-0.4444&	-0.4754\\
    CV($H$)&	-0.0318&	-0.0881&	-0.1256&	-0.1895&	-0.2466\\
    CV($K$)&	-0.2540&	-0.2354&	-0.1172&	-0.1166&	-0.2155\\
    \midrule
    Min($|H|$)&	-0.2149&	-0.3621&	\textbf{-0.5710}&	-0.2680&	-0.2552\\
    Min($|K|$)&	-0.1874&	-0.3180&	-0.2652&	-0.3457&	-0.1564\\
    Max($|H|$)&	-0.2181&	-0.3592&	-0.4120&	-0.2014&	-0.1078\\
    Max($|K|$)&	-0.3401&	-0.4051&	-0.4755&	-0.2916&	-0.1604\\
    Mean($|H|$)&	\textbf{-0.5937}&	\textbf{-0.5457}&	-0.4691&	-0.4211&	-0.4135\\
    Mean($|K|$)&	\textbf{-0.5979}&	\textbf{-0.5737}&	-0.4988&	-0.4148&	-0.4226\\
    SD($|H|$)&	\textbf{-0.6581}&	\textbf{-0.6288}&	\textbf{-0.5036}&	-0.3079&	-0.2546\\
    SD($|K|$)&	\textbf{-0.6274}&	\textbf{-0.6749}&	\textbf{-0.6416}&	-0.4598&	\textbf{-0.5031}\\
    CV($|H|$)&	-0.0187&	-0.0350&	0.0799&	0.2648&	0.2602\\
    CV($|K|$)&	-0.0193&	-0.0164&	-0.1332&	0.1310&	0.0743\\
    \bottomrule
    \end{tabularx}
    \caption{\textbf{Correlation coefficients $\rho_{X,Y}$ between the surface curvatures with different space separations and the average earthquake magnitude of subduction zones.} Note that 5.5km is the original space separation. Here, the average magnitude data is obtained based on the magnitude threshold $M_w \geq 6.0$. Coefficients with absolute value greater than 0.5 are in boldface.}
    \label{table:separation2}
\end{table*}

\begin{table*}[t]
    \centering
    \small
    \setlength{\tabcolsep}{4pt}
     \begin{tabularx}{\textwidth}{l|YYYYY}
    \toprule
    \multirow{2}{*}{\diagbox[width=2cm]{$X$}{$Y$}} & \multicolumn{5}{c}{\textbf{Number of earthquakes $n$ (with different space separations)}}\\
    \cmidrule(lr){2-6}
    & 5.5km & 16.5km & 27.5km & 38.5km & 49.5km\\ \midrule \midrule
    Min($H$)&	-0.4608&	-0.2928&	-0.0429&	-0.0604&	-0.0679\\
    Min($K$)&	-0.0277&	-0.0651&	-0.0941&	-0.0601&	0.0075\\
    Max($H$)&	0.1988&	0.1964&	0.1942&	0.2536&	0.2632\\
    Max($K$)&	0.2580&	0.1863&	0.0631&	0.2157&	0.0704\\
    Mean($H$)&	0.1213&	0.0571&	0.0071&	0.0189&	0.0684\\
    Mean($K$)&	0.3822&	0.3993&	0.4495&	0.4391&	0.4359\\
    SD($H$)&	-0.2455&	-0.2438&	-0.1586&	-0.0388&	0.0227\\
    SD($K$)&	-0.1369&	-0.2141&	-0.1923&	-0.0419&	-0.1224\\
    CV($H$)&	-0.0099&	-0.0467&	-0.0687&	-0.1757&	-0.2344\\
    CV($K$)&	-0.0394&	-0.3509&	-0.0997&	-0.0120&	0.1266\\
    \midrule
    Min($|H|$)&	\textbf{-0.5260}&	-0.3903&	-0.3097&	-0.2548&	-0.2845\\
    Min($|K|$)&	\textbf{-0.5140}&	\textbf{-0.5313}&	-0.1217&	-0.2830&	-0.4325\\
    Max($|H|$)&	0.4771&	0.3551&	0.1667&	0.2266&	0.1261\\
    Max($|K|$)&	0.1970&	0.1375&	0.0072&	0.0842&	-0.0965\\
    Mean($|H|$)&	-0.2226&	-0.1623&	-0.0955&	-0.0786&	-0.0966\\
    Mean($|K|$)&	-0.2758&	-0.3147&	-0.2924&	-0.2490&	-0.3339\\
    SD($|H|$)&	-0.2026&	-0.1741&	-0.0815&	0.0376&	0.0709\\
    SD($|K|$)&	-0.1462&	-0.2793&	-0.2836&	-0.1133&	-0.2190\\
    CV($|H|$)&	0.2908&	0.1807&	0.2163&	0.3462&	0.3698\\
    CV($|K|$)&	\textbf{0.5276}&	0.3724&	0.1298&	0.3711&	0.2175\\
    \bottomrule
    \end{tabularx}
    \caption{\textbf{Correlation coefficients $\rho_{X,Y}$ between the surface curvatures with different space separations and the earthquake productivity of subduction zones.} Note that 5.5km is the original space separation. Here, the earthquake numbers are obtained based on the magnitude threshold $M_w \geq 6.0$. Coefficients with absolute value greater than 0.5 are in boldface.}
    \label{table:separation3}
\end{table*}

\end{document}